\documentclass{article}

\usepackage{authblk}                    
\usepackage{graphicx,epsfig,epstopdf}   
\usepackage{booktabs}                   
\usepackage{amsmath,amsthm,amssymb}     
\usepackage{color}                      
\usepackage{cite}                       

\DeclareUnicodeCharacter{2212}{-}

\graphicspath{ {figures/} }

\usepackage[paperwidth = 17cm,          
paperheight = 24cm,
left = 1.75cm,
right = 1.75cm,
top = 2cm,
bottom = 2cm]{geometry}

\usepackage[colorlinks = true,          
linktocpage=true,
urlcolor = coolblack,
citecolor = red,
anchorcolor = blue,
linkcolor = blue]{hyperref}
\definecolor{coolblack}{rgb}{0.0, 0.18, 0.39}

\title{\textbf{Electroweak phase transition and gravitational waves in a two-component dark matter model}} 
\author[1]{Ahmad Mohamadnejad\thanks{mohamadnejad.a@lu.ac.ir}}
\affil[1]{\small Department of Physics, Lorestan University, Khorramabad, Iran}

\begin{document}
\maketitle

\begin{abstract}
We investigate an extension of the Standard Model (SM) with two candidates for dark matter (DM). One of them is a real scalar field and the other is an Abelian gauge field. Except for these two, there is another beyond SM field which has unit charge under a dark $ U_{D}(1) $ gauge symmetry. The model is classically scale invariant and the electroweak symmetry breaks because of the loop effects. Although SM is extended with a new dark symmetry and three fields, because of scale invariance, the parameter space is strictly restricted compared to other two-component DM models. We study both DM phenomenology and electroweak phase transition and show that there are some points in the parameter space of the model consistent with DM relic density and direct detection constraints, while at the same time can lead to first order electroweak phase transition. The gravitational waves produced during the phase transition could be probed by future space-based interferometers such as Laser Interferometer Space Antenna (LISA) and Big Bang Observer (BBO).
\end{abstract}

\newpage
\noindent\hrulefill
\tableofcontents
\noindent\hrulefill

\numberwithin{equation}{section}

\section{Introduction} \label{sec1}
The standard model (SM) is the most successful model in describing the observed phenomena, however, there are still
intriguing questions, such as the
nature and the origin of dark matter (DM), waiting to be answered.
There is a wide variety of astrophysical and
cosmological observations as well as theoretical
arguments that led the scientific community to adopt DM as an essential part of the standard
cosmological model (for a fascinating review on history of DM see \cite{Bertone:2016nfn}).

The evidence of DM was strong enough that many strategies have been pursued to reveal its particle nature. For decades, the leading theory of particle DM was a single-component thermal relic
with weak size couplings and mass, known as a
weakly interacting massive particle (WIMP). 
These one-component scenarios are increasingly constrained by experimental measurements. Therefore,
workers in the field compelled to examine more complex
models of dark sector including the multicomponent ones where the total relic abundance of DM is due to the existence of multiple DM species \cite{Zurek:2008qg,Profumo:2009tb,Aoki:2012ub,Biswas:2013nn,Gu:2013iy,Aoki:2013gzs,Kajiyama:2013rla,Bian:2013wna,Bhattacharya:2013hva,Geng:2013nda,Esch:2014jpa,Dienes:2014via,Bian:2014cja,Geng:2014dea,DiFranzo:2016uzc,Aoki:2016glu,DuttaBanik:2016jzv,Pandey:2017quk,Borah:2017xgm,Herrero-Garcia:2017vrl,Ahmed:2017dbb,PeymanZakeri:2018zaa,Aoki:2018gjf,Chakraborti:2018lso,Bernal:2018aon,Poulin:2018kap,Herrero-Garcia:2018qnz,YaserAyazi:2018lrv,Elahi:2019jeo,Borah:2019epq,Borah:2019aeq,Bhattacharya:2019fgs,Biswas:2019ygr,Nanda:2019nqy,Yaguna:2019cvp,Belanger:2020hyh,VanDong:2020bkg,Khalil:2020syr,DuttaBanik:2020jrj,Hernandez-Sanchez:2020aop,Chakrabarty:2021kmr,Yaguna:2021vhb,DiazSaez:2021pmg,DiazSaez:2021pfw}. After all, it should not be surprising if the dark sector has
multiple species like SM itself.

On the other hand, SM has a crossover rather than a true phase transition \cite{Kajantie:1996mn}, while
some extensions of the SM, e.g., with DM candidates \cite{Chala:2016ykx,Soni:2016yes,Flauger:2017ged,Baldes:2017rcu,Chao:2017vrq,Beniwal:2017eik,Huang:2017rzf,Huang:2017kzu,Madge:2018gfl,Bian:2018mkl,Bian:2018bxr,Shajiee:2018jdq,Kannike:2019wsn,YaserAyazi:2019caf,Mohamadnejad:2019vzg,Kannike:2019mzk,Paul:2019pgt,Barman:2019oda,Marfatia:2020bcs,Alanne:2020jwx,Han:2020ekm,Wang:2020wrk,Deng:2020dnf,Chao:2020adk,Zhang:2021alu,Liu:2021mhn}
, lead to first order phase transitions with gravitational wave (GW) signals.
In the case of first order phase transition, just below the
critical temperature, the Universe goes from a metastable quasi-equilibrium state
into a stable equilibrium state, through a process of bubble nucleation, growth, and merger which generates GWs
\cite{Witten:1980ez,Guth:1981uk,Steinhardt:1980wx,Steinhardt:1981ct,Witten:1984rs}.
GW signatures are therefore
a new window towards new physics, complementary to that provided by the
Large Hadron Collider.
Another motivation for studying electroweak phase transitions is the requirements to explain the matter-antimatter asymmetry in the universe \cite{Shaposhnikov:1987tw}, which one of them,  departure
from thermal equilibrium, is inevitable in a first-order phase transition.

The first direct detection of GWs was performed in 2015 \cite{LIGOScientific:2016aoc}. The signal came from the strongest astrophysical sources of GWs, i.e., compact binary systems, in frequency from 35 to 250 Hz.
Here, we are interested in production of
GWs by first-order phase transitions (for a recent review see \cite{Caprini:2019egz}). For GWs sourced by cosmological phase transitions, the relevant mission is Laser
Interferometer Space Antenna (LISA) \cite{LISA:2017pwj} which is a space-based interferometric gravitational
wave detector working with three satellites orbiting the Earth.
LISA is most sensitive at frequencies in the range $ 10^{-3} - 10^{-2} $ Hz and its planned launch year is 2034 with a mission life-time of 4 years.
The Big Bang Observer (BBO) \cite{Crowder:2005nr} is a proposed follow-up experiment
consisting of four LISA-like detectors.

In this paper, a model with three motives will be presented. We come up with a beyond SM model to provide a solution for DM problem, hierarchy problem, and vacuum instability. 
As a solution for hierarchy problem \cite{Bardeen}, the model has classical conformal symmetry. On the other hand, SM suffers from vacuum instability and beyond SM models with bosonic degrees of freedom can probably solve this issue. As mentioned, due to the strict constraints of direct detection on one-component DM models, two component DM models are more appropriate. Therefore, our two-component DM model consists (bosonic) scalar and vector DM. We assume that the dark sector
interacts with the SM particles only through the Higgs portal. The dark sector consists of three new fields, a real scalar, a complex singlet and a real vector field and it introduces a dark $ U_{D}(1) $ gauge symmetry. Our model is a two-component DM model which both DM particles are bosons: one DM particle is spin zero and the other is spin one.
As a potential solution to the hierarchy problem, we constrain our model to be a classically scale-invariant extension of the SM.
Within this framework, all the particle
masses are generated
dynamically, by means of the Coleman-Weinberg mecha.nism \cite{Coleman:1973jx}. After constructing the model, we study DM phenomenology including relic density and DM-Nucleon cross section.
DM relic density is reported by Planck Collaboration \cite{Planck:2018vyg}, and DM-Nucleon cross section is constrained by direct detection experiments
such as the LUX \cite{LUX:2016ggv}, PandaX-II \cite{PandaX-II:2016vec} and XENON1T \cite{XENON:2018voc}. These experiments are gradually approaching the so-called neutrino floor which is
 the ultimate sensitivity of future direct
detection experiments \cite{Billard:2013qya}.
We also concentrate on investigating
the possibility of achieving a strongly first-order electroweak phase transition within the parameter space of the model. 
To study electroweak phase transition, we present
the complete expression of the finite-temperature 1-loop
effective potential, including the contributions of the
resummed thermal bosonic daisy diagrams, and show
that the finite-temperature corrections induce a first-order
electroweak phase transition. We identify regions of parameter space of the model which is consistent with DM relic density and direct detection constraints, while
simultaneously realizing a strongly first-order electroweak
phase transition.
The GW signal from the phase transition is sufficiently strong to be detectable by LISA and BBO.

The rest of the paper is organized as follows. In section \ref{sec2}, we introduce our model. Section \ref{sec3} is dedicated to DM phenomenological constraints. In section \ref{sec4}, we study the electroweak phase transition and
GWs spectrum. Our result is given in section \ref{sec5} where we simultaneously consider DM phenomenolgy and GW spectrum using two benchmark points (BMs) as shown in table \ref{table1}.
Finally, we conclude in
section \ref{sec6}.

\section{The model} \label{sec2}
Our model consists three beyond SM fields which all of them are bosons, namely, a complex singlet $ \phi $, a real singlet $ S $, and a real Abelian vector field $ V_{\mu} $. In our setup, the complex scalar field $ \phi $ has unit charge under a dark $ U_{D}(1) $ gauge symmetry with the vector field $ V_{\mu} $. All of these fields are singlet under SM gauge group. However, the dark sector is invariant under
the charge conjugation of $ U_{D}(1) $ and parity of $ S $:

\begin{equation} \label{2-1}
\phi \rightarrow \phi^{*},  \, \quad V_{\mu} \rightarrow - V_{\mu}, \quad \text{and} \quad S \rightarrow -S .
\end{equation}

Due these two symmetries, the model can have two component DM. In the dark sector the discrete symmetry $ V_{\mu} \rightarrow - V_{\mu} $ forbids the kinetic mixing between the SM $ U_{Y}(1) $ gauge boson $ B_{\mu} $ and the vector field $ V_{\mu} $ which makes $ V_{\mu} $ stable and a DM candidate. The other DM candidate is due to $ S \rightarrow -S $ symmetry which makes the real singlet field stable.

The Lagrangian is given by

\begin{equation} \label{2-2}
 {\cal L} ={\cal L}_{\text{SM}} + \frac{1}{2}(\partial_{\mu} S)(\partial^{\mu} S) + (D_{\mu} \phi)^{*} (D^{\mu} \phi) - \frac{1}{4} V_{\mu \nu} V^{\mu \nu} - V(H,\phi,S) , 
\end{equation}
where $ V_{\mu \nu} = \partial_{\mu} V_{\nu} - \partial_{\nu} V_{\mu}  $, $ D_{\mu} \phi = (\partial_{\mu} + i g V_{\mu}) \phi $, and $ {\cal L}_{\text{SM}} $ is the SM Lagrangian without the Higgs potential term. 

We constrain $ V(H,\phi,S) $ by:
\begin{itemize}
\item gauge symmetry,
\item $ Z_{2} $ symmetry,
\item scale invariance, and
\item renormalizablity.
\end{itemize}
Regarding these constraints $ V(H,\phi,S) $ reads

\begin{align} \label{2-3}
V(H,\phi,S) = & \, \lambda_{H} (H^{\dagger}H)^{2} + \lambda_{\phi} (\phi^{*}\phi)^{2}  
+ \lambda_{H \phi} (H^{\dagger}H)(\phi^{*}\phi) \nonumber \\
&+ \frac{1}{2} \lambda_{H S} (H^{\dagger}H)S^2
+ \frac{1}{2} \lambda_{\phi S} (\phi^{*}\phi)S^2 + \frac{1}{4} \lambda_{S} S^4.
\end{align}

In this model, dark sector interacts with SM via Higgs portal. Because of $ Z_{2} $ symmetry, the real singlet field, $ S $, does not get vacuum expectation value (VEV), but electroweak symmetry, as well as Abelian dark symmetry spontaneously break after
$ H $ and $ \phi $ develop VEVs. In the unitary gauge, $ H^{\dagger} = (0,\frac{h_{1}}{\sqrt{2}}) $ and $ \phi = \frac{h_{2}}{\sqrt{2}} $, the tree level potential is given by
\begin{align} \label{2-4}
V_{tree}(h_1,h_{2},S) = & \, \frac{1}{4} \lambda_{H} h_{1}^{4} + \frac{1}{4} \lambda_{\phi} h_{2}^{4} 
+ \frac{1}{4} \lambda_{H \phi} h_{1}^{2} h_{2}^{2} \nonumber \\
&+ \frac{1}{4} \lambda_{H S} h_{1}^{2}S^2
+ \frac{1}{4} \lambda_{\phi S} h_{2}^{2}S^2 + \frac{1}{4} \lambda_{S} S^4.
\end{align}
The local minimum of the tree level potential defines the VEVs of the fields which we take as $ (\langle h_{1} \rangle,\langle h_{2} \rangle,\langle S \rangle) = (\nu_1,\nu_2,0) $, leading to the following condition 
\begin{align} \label{2-5}
\lambda _H>0\land \lambda _{H\phi}<0\land \lambda _{\phi }=\frac{\lambda _{H \phi}^2}{4 \lambda _H}\land
\frac{\nu _1}{\nu _2} = \sqrt{-\frac{\lambda _{H \phi}}{2\lambda _H}} .
\end{align}
The last relation defines the flat direction in field space where the tree level potential is minimum, along this direction
 $ V_{tree}(\nu_1,\nu_{2},0) = 0 $.
Now we substitute $ h_{1} \rightarrow  \nu_1 + h_{1} $ and $ h_{2} \rightarrow  \nu_2 + h_{2} $ which mixes $ h_{1} $ and $ h_{2} $.
The mass eigenstates, $ h $ and $ \varphi $, can be obtained from the following rotation
\begin{equation} \label{2-6}
\begin{pmatrix}
h\\\varphi\end{pmatrix}
 =\begin{pmatrix}  cos \alpha~~~  -sin \alpha \\sin \alpha  ~~~~~cos \alpha
 \end{pmatrix}\begin{pmatrix}
h_1 \\  h_{2}
\end{pmatrix},
\end{equation}
where $ \tan \alpha = \nu_1/\nu_2 $ ($ \langle h \rangle = 0 $ and $ \langle \varphi \rangle = \nu = \sqrt{\nu_1^2+\nu_2^2} $).
We identify $ h $, which is perpendicular to the flat direction, as the SM-like Higgs observed at the LHC with $ M_{h}=125 $ GeV \cite{ATLAS:2012yve,CMS:2012qbp}. On the other hand, we know from SM that $ \nu_1 = 246 $ GeV.
The field $ \varphi $ is along the flat direction, thus its tree level mass is zero. However, 1-loop correction leads to a specific value along flat direction as the minimum of the potential which gives the following mass to $ \varphi $ (see subsection~\ref{4.1}):
\begin{equation}
M_{\varphi}^{2} = \frac{1}{8 \pi^{2} \nu^{2}} \left( M_{h}^{4} + M_{S}^{4}+ 3  M_{V}^{4} + 6  M_{W}^{4} + 3  M_{Z}^{4} - 12 M_{t}^{4}   \right) , \label{2-7}
\end{equation}
where $ M_{S,V,W,Z,t} $ being the masses for scalar DM, vector DM, W and Z gauge bosons, and top quark, respectively, after symmetry breaking.
We can substitute the parameters of the Lagrangian using
\begin{align}\label{2-8}
&\lambda _{H\phi}= -\frac{M_h^2}{\nu ^2},\lambda _{\phi }= \frac{M_h^2}{2\nu ^2} \tan^2\alpha ,\lambda _H= \frac{M_h^2}{2\nu ^2} \cot^2\alpha , \nonumber \\
&M_V=g \cos \alpha \, \nu, M_S^2\text{ = }\frac{1}{2}  \left( \lambda_{\phi S} \cos ^2 \alpha +\lambda _{HS} \sin ^2 \alpha  \right) \nu ^2.
\end{align}
According to these relations there are five free parameters which we choose them as $ M_S, M_V, g, \lambda _{\phi S}$, and $ \lambda _{S} $. On the other hand, $ \lambda _{S} $ is irrelevant in DM phenomenology and phase transition studied in this paper, therefore we left with only four parameters.

\section{Dark matter phenomenology} \label{sec3}
In the following we will focus on the DM relic density constraint
reported by Planck collaboration \cite{Planck:2018vyg} and the available data on direct DM detection.

\subsection{Relic density} \label{3.1}
\begin{figure}[!htb]
\begin{center}
\includegraphics[scale=1.0]{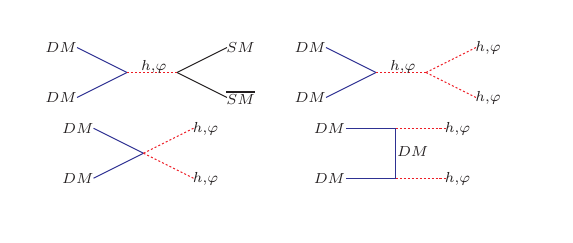}
\end{center}
\vspace{-1.0cm}
\caption{Pair annihilation processes for both scalar and vectorial DM components. \label{Annihilation}}
\end{figure}

\begin{figure}[!htb]
\begin{center}
\includegraphics[scale=1.0]{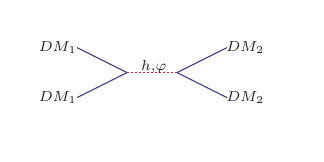}
\end{center}
\vspace{-1.0cm}
\caption{DM conversion via Higgs portal. \label{Conversion}}
\end{figure}

It is considered in general that in WIMP scenarios, the DM relic density is inversely proportional to the thermally averaged DM annihilation cross section into SM particles. In the case of two-component DM, the situation is more interesting since there are additional important processes such as conversion of one DM component into another which complicates the analysis.
We use the public numerical code {\tt micrOMEGAs} \cite{Barducci:2016pcb} for solving the two Boltzmann equations governing
the cosmological evolution of our DM candidates, the scalar DM ($ S $) and the Vector DM ($ V $).
The Boltzmann equations are determined by three types of processes:
\begin{itemize}
\item
DM Annihilation: pair annihilation of both DM components into SM particles,
\item
DM Conversion: which converts one DM component into another, and
\item
DM Semi-(co)annihilation: which could change
the abundances of both DM components
\end{itemize}
The relevant diagrams for the annihilation processes of the two DM components are presented in figure \ref{Annihilation}. There are no (subleading) semi-annihilation and co-annihilation processes in our scenario. As we see in figure \ref{Annihilation}, Feynman diagrams are the same for both DM components. In these figure, $ SM $ and $ \overline{SM} $ stand for massive SM particles and anti-particles, respectively.
Besides DM annihilation into SM particles, the two
DM candidate can also annihilate into each other (DM conversion: $ S \, \, S \longleftrightarrow V \, \, V $) which are shown in figure \ref{Conversion}.

The coupled Boltzmann equations for scalar $S$ and vector $V$ DM are given by:
\begin{equation}
\frac{d n_{V}}{dt}+3Hn_{V}=-\sum_{j} \langle\sigma_{VV\rightarrow jj}\upsilon\rangle (n^2_{V}-n^2_{V,eq})-\langle\sigma_{VV\rightarrow SS}\upsilon\rangle (n^2_{V}-n^2_{V,eq}\frac{n^2_{S}}{n^2_{S,eq}})
\label{3-1},
\end{equation}
\begin{equation}
\frac{d n_{S}}{dt}+3Hn_{S}=-\sum_{j} \langle\sigma_{SS\rightarrow jj}\upsilon\rangle (n^2_{S}-n^2_{S,eq})-\langle\sigma_{SS\rightarrow VV}\upsilon\rangle (n^2_{S}-n^2_{S,eq}\frac{n^2_{V}}{n^2_{V,eq}})
\label{3-2},
\end{equation}
where $j$ runs over SM massive particles and $ h,\varphi $.
By changing the variable, $x = M/T$ and $Y=n/s$, 
where $T$ is the photon temperature and $ s $ is the entropy density, one can rewrite the Boltzmann equations in terms of $Y=n/s$:
\begin{align}
\frac{dY_{V}}{dx}=&-\sqrt{\frac{45}{\pi}} M_{pl} \, g_{*}^{1/2} \, \frac{M}{x^{2}}
[\sum_{j}\langle\sigma_{V V \rightarrow jj}v\rangle(Y_{V}^{2}-Y_{V,eq}^{2}) \nonumber \\
&+\langle\sigma_{V V \rightarrow SS}v\rangle(Y_{V}^{2}-Y_{V,eq}^{2} \frac{Y_{S}^{2}}{Y_{S,eq}^{2}})], \label{3-3} 
\end{align}
\begin{align}
\frac{dY_{S}}{dx} =& - \sqrt{\frac{45}{\pi}} M_{pl} \, g_{*}^{1/2} \, \frac{M}{x^{2}}
[\sum_{j}\langle\sigma_{SS \rightarrow jj }v\rangle(Y_{S}^{2}-Y_{S,eq}^{2}) \nonumber \\
&+\langle\sigma_{SS \rightarrow V V}v\rangle(Y_{S}^{2}-Y_{S,eq}^{2} \frac{Y_{V}^{2}}{Y_{V,eq}^{2}})], \label{3-4}
\end{align}
where $g_{*}^{1/2}$ is the degrees of freedom parameter and $M_{pl}$ is the Planck mass. The last terms in summations are new terms in Boltzmann equations which describe the conversion of DM particles into each other. 
Since these two cross sections are described by the same matrix element, we expect $ \langle\sigma_{V V \rightarrow SS}v \rangle$ and $\langle \sigma_{SS \rightarrow V V}v\rangle $ are not independent and their relation is:
\begin{equation} \label{3-5}
Y_{V,eq}^{2} \langle\sigma _{V V \rightarrow SS}v\rangle =
Y_{S,eq}^{2} \langle\sigma _{SS \rightarrow V V}v\rangle .
\end{equation}

The interactions between the two DM candidates take place by exchanging two scalar mass eigenstates $h$ and $\varphi$ where the coupling of $ V $ to $ h $ is suppressed by $ sin \, \alpha $. Therefore,
it is usually the $ \varphi $-mediated diagram that gives the dominant contribution. Notice that the conversion of the heavier particle into the lighter one is relevant.
The relic density for each DM candidate is related to $ Y $ at the present temperature through
$ \Omega_{S,V} h^2 = 2.755 \times 10^8 \frac{M_{S,V}}{GeV} Y_{S,V} (T_0) $, where $ h $ is the Hubble expansion rate at present times in units of 100 $ (km/s)/Mpc $, and the total relic density of DM according to the data by the Planck collaboration should be \cite{Planck:2018vyg},

\begin{equation} \label{3-6}
 \Omega_{\rm DM} h^2 = \Omega_{S} h^2 + \Omega_{V} h^2 = 0.120 \pm 0.001 .
\end{equation}
Finally, we define the fraction of the DM density
of each component by,
\begin{equation} \label{3-7}
 \xi_{V} = \frac{\Omega_{V}}{\Omega_{DM}}, \quad
  \xi_{S} = \frac{\Omega_{S}}{\Omega_{DM}}, \quad \rm with \, \, \, \xi_{V}  + \xi_{S}  = 1.
\end{equation}

\begin{figure}[!htb]
\begin{center}
\centerline{\hspace{0cm}\epsfig{figure=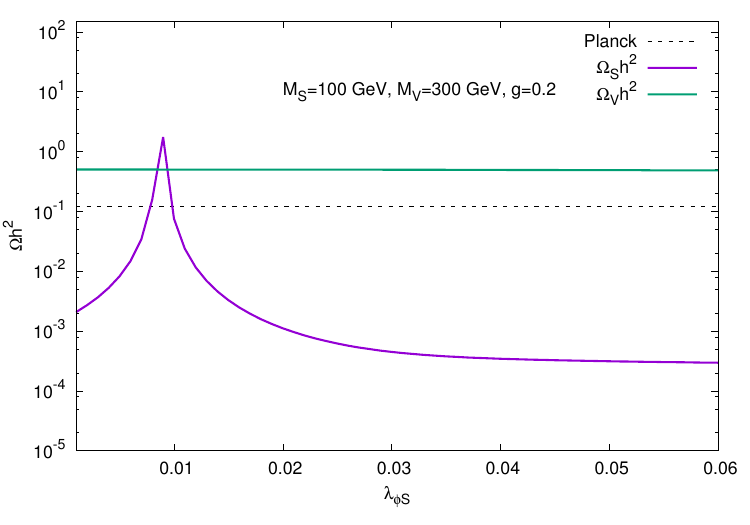,width=6cm}\hspace{0.3cm}\epsfig{figure=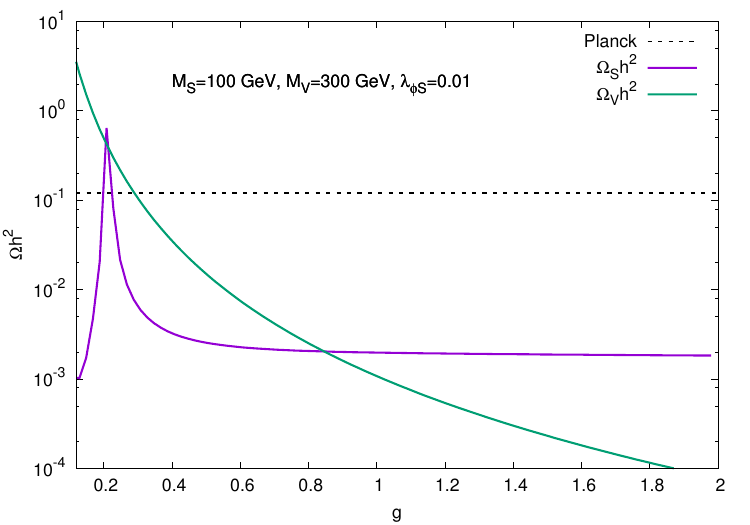,width=6cm}}
\centerline{\vspace{0.5cm}\hspace{0.5cm}(a)\hspace{6cm}(b)}
\centerline{\hspace{0cm}\epsfig{figure=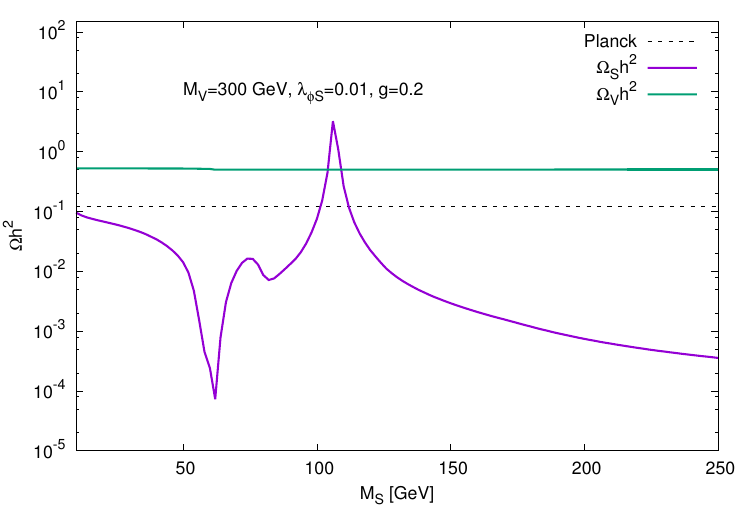,width=6cm}\hspace{0.3cm}\epsfig{figure=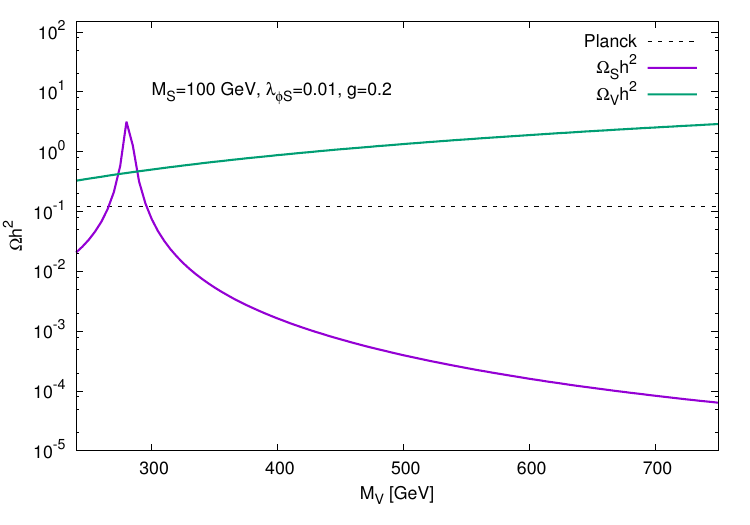,width=6cm}}
\centerline{\vspace{-1.2cm}\hspace{0.5cm}(c)\hspace{6cm}(d)}
\centerline{\vspace{-0.0cm}}
\end{center}
\caption{Variation of DM Relic density respect to parameter space}\label{Relic}
\end{figure}

Figure \ref{Relic} depicts the relic density of both DM components as a function of free parameters of the model.
According to these figure, $ \lambda_{\phi S} $, and $ M_S $ are not relevant in $ \Omega_{V} $ (Note that, in general, $ \Omega_{V} $ depends on $ M_S $, because $ M_{\varphi} $ is a function of $ M_S $). However, $ \Omega_{S} $ depends on all four parameters. In this case, besides  $ \lambda_{\phi S} $ and $ M_S $ which are relevant, $ \lambda_{H S} $ can also change $ \Omega_{S} $. On the other hand, considering eq.~(\ref{2-8}), $ \lambda_{H S} $ is dependent in all free parameters:
\begin{equation} \label{3-8}
\lambda_{H S}=\frac{2 M_S^2/\nu^2 - \lambda_{\phi S} \cos^2 \alpha}{\sin^2 \alpha}=\frac{1}{\nu_{1}^2} \left( 2 M_S^2 - \frac{\lambda_{\phi S}}{g^2} M_V^2 \right) .
\end{equation}
So it is not a surprise that all free parameters can effect the relic density of scalar DM. The minimum in figure \ref{Relic} (c) at $ M_S \simeq \frac{M_h}{2} $ is due to resonance, and maximum in all four diagrams is due to $ \lambda_{H S} \simeq 0 $.

\subsection{Direct detection} \label{3.2}

The direct detection experiments aim to study DM-Nucleon interactions.
These events induced by DM particles from the Milky Way’s halo.
The Standard Halo Model assumes that the DM particles are distributed in an isotropic isothermal sphere with a Maxwellian velocity distribution.
The local DM density $ \rho_0 $ adopted for the interpretation of direct detection experiments is $ \rho_0 =0.3 \, GeV/c^2 / cm^3 $.
The possibility of DM direct detection in the form of WIMPs was first discussed in \cite{Goodman:1984dc}. The idea is simple: since in
most scenarios the WIMP carries no electric charge, therefore it will not interact with the atomic electrons, however, DM particles can elastically scatter off the atomic nucleus and the momentum transfer gives rise to a nuclear recoil which might be detectable.

\begin{figure}[!htb]
\begin{center}
\includegraphics[scale=1.2]{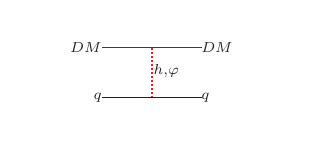}
\end{center}
\vspace{-1.3cm}
\caption{DM-quark interaction via t channel exchanges of the $ h $ and $ \varphi $ states. \label{DirectFeyn}}
\end{figure}

In our model both DM candidates interact with quarks via Higgs portal, see figure \ref{DirectFeyn}, which results in a spin independent DM-Nucleon cross section.

The relevant DM-quark interaction terms in Lagrangian are:
\begin{align} \label{3-9}
\mathcal{L}_q &= - \sum_q \frac{m_q}{\nu_1} \cos \alpha \, h \, \overline{q} q + \frac{m_q}{\nu_1} \sin \alpha \, \phi \, \overline{q} q , \nonumber \\
\mathcal{L}_S &= \lambda_{hss} h S^2 + \lambda_{\phi ss} \phi S^2  , \nonumber \\
\mathcal{L}_V &= \lambda_{hvv} h V_{\mu}V^{\mu} + \lambda_{\phi vv} \phi V_{\mu}V^{\mu},
\end{align}
where q stands for quarks and 
\begin{align} \label{3-10}
\lambda_{hss} &= - \frac{1}{2} \left(\nu _1 \cos \alpha  \lambda _{HS}-\nu _2 \sin \alpha  \lambda_{\phi S}\right), \nonumber \\
\lambda_{\phi ss} &= - \frac{1}{2} \left(\nu _2 \cos \alpha  \lambda_{\phi S}+\nu _1 \sin \alpha  \lambda _{{HS}}\right) , \nonumber \\
\lambda_{hvv} &=  -\sin \alpha \,  g^2 \nu_2, \nonumber \\
\lambda_{\phi vv} &=  + \cos \alpha  \, g^2 \nu_2.
\end{align}

DM scattering on nuclei have a characteristic energy scale of the order of 1 GeV and very low momentum exchange between the DM and the nucleon. On the other hand, the low DM velocity
allows to consider this process in the non{relativistic limit. Therefore, the scattering of
DM with nucleons can be described by effective
four field interactions between the DM and the SM quarks.
After integrating the scalar mediators out, the low-energy 5-dimensional effective interaction of the DM with quarks will be
\begin{align} \label{3-11}
\mathcal{L}_{S-q} &= \alpha_s S^2 \sum_q m_q \overline{q} q , \nonumber \\
\mathcal{L}_{V-q} &=  \alpha_v V_{\mu}V^{\mu} \sum_q m_q \overline{q} q,
\end{align}
where
\begin{align} \label{3-12}
\alpha_s &= \frac{\lambda_{HS}}{2} \left(\frac{\cos ^2\alpha }{M_h^2}+\frac{\sin ^2\alpha }{M_{\varphi }^2}\right)
-\frac{\lambda_{\phi S} \cos^2 \alpha}{2}\left( \frac{1}{M_{h}^2}-\frac{1}{M_{\varphi}^2} \right) , \nonumber \\
\alpha_v &= g^2 \cos^2 \alpha\left( \frac{1}{M_{h}^2}-\frac{1}{M_{\varphi}^2} \right) .
\end{align}

From this, it is possible to obtain effective interactions between the
DM particle and a nucleon which gives DM-Nucleon cross section of scalar DM and vector DM \cite{Kanemura:2010sh}
\begin{align} \label{3-13}
\sigma_s &= \alpha_s^2 \frac{M_N^4}{\pi(M_N+M_S)^2} f_N^2   , \nonumber \\
\sigma_v &= \alpha_v^2 \frac{M_N^4}{\pi(M_N+M_V)^2} f_N^2 ,
\end{align}
where $ M_N $ is the nucleon mass and $ f_N=0.3 $ parametrizes the Higgs-Nucleon coupling.

The Higgs portal scattering that we discussed here led to spin independent interactions of the DM with nuclei. Some of the present constraints on DM-Nucleon spin independent interactions come from the world
leader experiments, such as LUX \cite{LUX:2016ggv}, PandaX-II \cite{PandaX-II:2016vec} and XENON1T \cite{XENON:2018voc}. 
Finally, the DARWIN experiment \cite{DARWIN:2016hyl}, with sensitivity close to the irreducible background coming from scattering of SM neutrinos on nucleons (the so-called neutrino floor \cite{Billard:2013qya}), would be the ultimate DM detector.
In Section \ref{sec5}, we constrain the
model with the results of the PandaX-II experiment \cite{PandaX-II:2016vec} which set an upper limit on the spin-independent WIMP-Nucleon cross section with the lowest exclusion at $ M_{DM} = 40 $ GeV:
\begin{equation} \label{3-14}
\text{PandaX-II}: \quad \sigma_{\text{DM-N}} \lesssim 8.6 \times 10^{-47} \, \text{cm}^3 .
\end{equation}
Note that in above constraint, it is assumed that the local DM density is only provided by one DM specie. However, in our scenario both scalar and vector DM contribute to the local DM
density. Assuming that the contribution of each DM
candidates to the local DM density is the same as their contribution to
the relic density, many authors constrain the rescaled DM-Nucleon cross sections, i.e., $ \xi_{S} \sigma_s $ and $ \xi_{V} \sigma_v $, with experimental results. However, both DM candidates contribute the DM signatures and one must combine both signatures. For the large DM mass case, where DM energy is much larger than detector threshold energy, the statistical combination is easy and the direct detection constraint reads \cite{Hur:2007ur}
\begin{equation} \label{3-15}
 \xi_S \frac{\sigma_s}{M_S} + \xi_V \frac{\sigma_v}{M_V} \lesssim \frac{\sigma}{M} \bigg\rvert_{\text{PandaX-II}},
\end{equation}
where
\begin{equation} \label{3-16}
\frac{\sigma}{M} \bigg\rvert_{\text{PandaX-II}} \simeq 0.001 \, \frac{\text{zb}}{\text{GeV}},
\end{equation}
for $ M \gtrsim 40 $ GeV.

\begin{figure}[!htb]
\begin{center}
\centerline{\hspace{0cm}\epsfig{figure=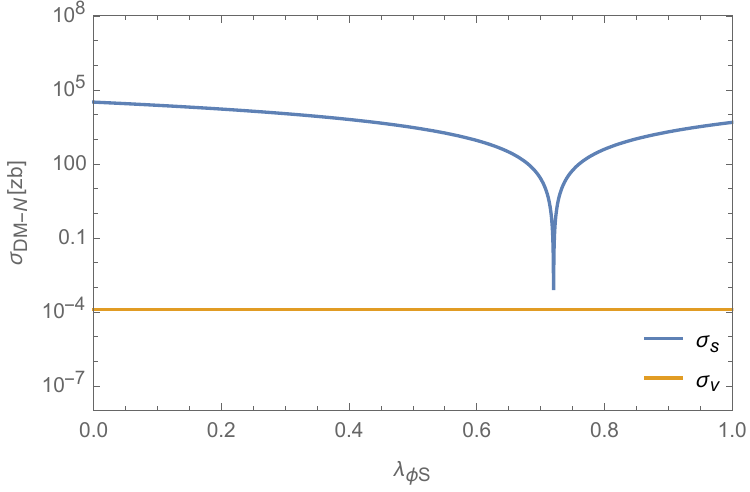,width=6.0cm}\hspace{0.3cm}\epsfig{figure=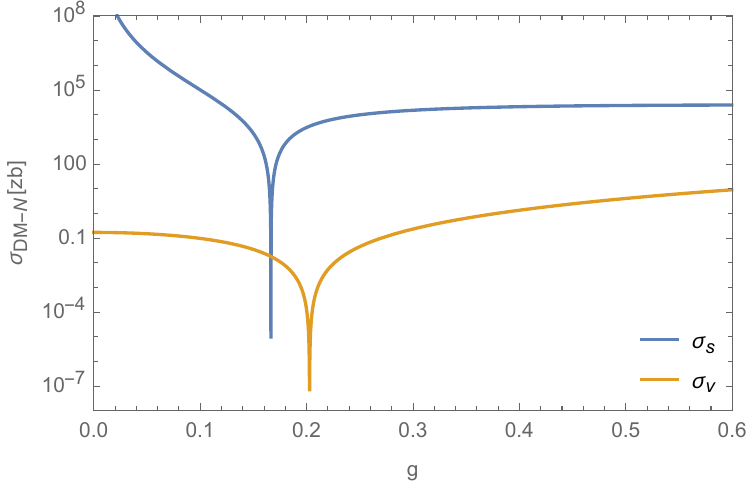,width=6.0cm}}
\centerline{\vspace{0.5cm}\hspace{0.5cm}(a)\hspace{6cm}(b)}
\centerline{\hspace{0cm}\epsfig{figure=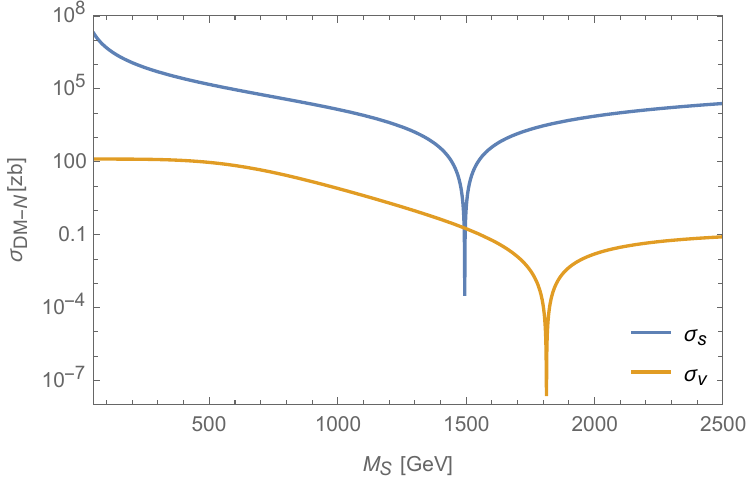,width=6.0cm}\hspace{0.3cm}\epsfig{figure=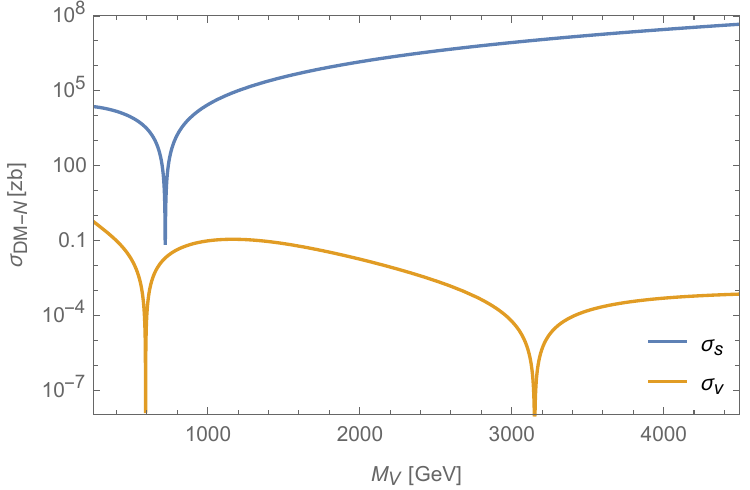,width=6.0cm}}
\centerline{\vspace{-1.2cm}\hspace{0.5cm}(c)\hspace{6cm}(d)}
\centerline{\vspace{-0.0cm}}
\end{center}
\caption{Variation of DM-Nucleon spin-independent cross section respect to parameter space. In all digrams the fixed parameters are: $ M_S = 1800 \, \text{GeV} \, , \, M_V=600 \, \text{GeV} \, , \, g= 0.2 \, , \, \lambda_{\phi S} = 0.5$. }\label{Direct}
\end{figure}

We have depicted DM-Nucleon cross section as a function of free parameters of the model in figure \ref{Direct}. Again, $ \lambda_{\phi S} $ is irrelevant in vector DM phenomenology, e.g., $ \sigma_v $. 
In all diagrams there are some dips where $ \alpha_s $ or $ \alpha_v $ vanishes. According to eq.~(\ref{3-12}), If $ \lambda_{HS}=0 $, then we expect that dips occur at the same place where $ M_h \simeq M_{\varphi} $. But, in general , the dips occur at different places as we see in figure \ref{Direct}. Another interesting feature is the double dips of $ \sigma_v $ in (d) diagram of figure \ref{Direct}. If we solve $ \alpha_v=0 \, (\Rightarrow M_h = M_{\varphi}) $ to find $ M_V $, we will find two solutions corresponding these double dips.

In our model, indirect detection limits are not competitive with the ones from direct detection and we will not explicitly discuss them here.

\section{Electroweak phase transition} \label{sec4}
In order to study the electroweak phase transition in our model, we need to construct the effective potential. In our scenario, the effective potential is a function of the scalar field $ \varphi $ and temperature $ T $. As the Universe cools down, the VEV varies from $ \langle \varphi \rangle = 0 $ to $ \langle \varphi \rangle = \nu \neq 0 $. In the following subsections we derive 1-loop effective potential and study GWs during the cosmological phase transition.

\subsection{One-loop effective potential} \label{4.1}
The effective potential were initially studied at 1-loop level by Coleman and Weinberg \cite{Coleman:1973jx}. A few years later, Gildener and Weinberg, presented their formulation for a scale invariant theory with many scalar fields \cite{Gildener:1976ih}. As we discussed in Sec.~\ref{sec2}, along the flat direction, the tree-level potential is zero, therefore, 1-loop corrections are dominated. The 1-loop effective potential at zero temperature is given by
\begin{equation}
V^{1-loop} (\varphi) = a \varphi^{4} + b \varphi^{4} \ln \frac{\varphi^{2}}{\Lambda^{2}} , \label{4-1}
\end{equation}
where
\begin{align}
& a =  \frac{1}{64 \pi^{2} \nu^{4}}  \sum_{k=1}^{n} g_{k}  M_{k}^{4} \left(  \ln \frac{M_{k}^{2}}{\nu^{2}} - C_{k}  \right)   , \nonumber \\
& b = \frac{1}{64 \pi^{2} \nu^{4}} \sum_{k=1}^{n} g_{k}  M_{k}^{4} , \label{4-2}
\end{align}
and $ \Lambda $ is the renormalization group (RG) scale. The other parameters are: $ C_{k}=3/2 $ ($ 5/6 $) for scalars/spinors (vectors), $ M_{k} $ for the measured mass of particles, and $ g_{k} $ for the number of degrees of
freedom of the particle k (it is positive for bosons and negative for fermions). In order to have a non-zero VEV, the potential~(\ref{4-1}) should have a minimum at $ \varphi \neq 0 $:
\begin{align}
& \frac{d V^{1-loop}}{d \varphi} \bigg\rvert_{\langle \varphi \rangle \neq 0} = 0, \nonumber \\
& \frac{d^2 V^{1-loop}}{d \varphi^2} \bigg\rvert_{\langle \varphi \rangle \neq 0} > 0,  \label{4-3}
\end{align}
which leads to
\begin{equation}
\langle \varphi \rangle = \nu = \Lambda e^{-(\frac{a}{2b} + \frac{1}{4})} \quad \text{and} \quad b>0. \label{4-4}
\end{equation}
Considering both eq.~(\ref{4-1}) and eq.~(\ref{4-4}) one can substitute RG scale $ \Lambda $ and find a final expression for the 1-loop potential in terms of $ b $ coefficient and the true vacuum expectation value $ \nu $:
\begin{equation}
V^{1-loop} (\varphi) = b \varphi^{4} \left(  \ln \frac{\varphi^{2}}{\nu^{2}} - \frac{1}{2} \right) , \label{4-5}
\end{equation}
and for the mass of $ \varphi $ we have $ M_{\varphi}^2 = 8 b \nu^2 $ which gives eq.~(\ref{2-7}). Vacuum stability of a model depends on the behavior of the effective potential.
If the vacuum of the effective potential is a global minimum, then the vacuum is absolutely stable.
Vacuum stability up to Planck scale puts constraint on the parameters of models. 
After the discovery of 125 GeV Higgs boson at the LHC, we know that for SM the vacuum is not stable if no new physics is assumed (for a brief review of vacuum stability in SM see \cite{Tang:2013bz}).
In conformal models tree-level potential is zero along flat direction and 1-loop contribution determines the behavior of the effective potential.
According to conditions (\ref{4-3}), the effective potential (\ref{4-5}) has global minimum if $ b > 0 $. To fulfill this condition, considering eq.~(\ref{4-2}), we need new bosonic degrees of freedom with constrained masses. Albeit, for full treatment of vacuum stability, one should obtain one-loop $ \beta $-functions and solve renormalization group equations (RGEs) in order to derive running coupling constants (see, e.g., \cite{Karam:2015jta,Karam:2016rsz} where classically scale-invariant non-Abelian extensions of the SM are constructed satisfying perturbativity and stability up to the Planck scale.)

Apart from 1-loop zero-temperature potential~(\ref{4-5}) , the 1-loop corrections at finite temperature also contribute to the effective potential which is given by \cite{Dolan:1973qd}
\begin{equation}
V_{T\neq0}^{1-loop} (\varphi,T) = \frac{T^{4}}{2 \pi^{2}}  \sum_{k=1}^{n} g_{k} J_{\text{B,F}} \left( \frac{M_k}{\nu} \frac{\varphi}{T} \right) , \label{4-6}
\end{equation}
where $ J_{\text{B,F}} $ are thermal functions:
\begin{equation}
J_{\text{B,F}}(x) =  \int_{0}^{\infty} dy \, y^{2} \ln \left(1 \mp e^{- \sqrt{y^{2}+x^{2}}} \right). \label{4-7}
\end{equation}
We approximate thermal functions $ J_{\text{B}}(x) $ and $ J_{\text{F}}(x) $ in terms of modified Bessel functions of the second kind, $ K_2\left(x\right) $,
\begin{align}
&J_{\text{B}}(x) \simeq -\sum _{k=1}^3 \frac{1}{k^2} x^2 K_2\left(k x\right), \nonumber \\
&J_{\text{F}}(x) \simeq -\sum _{k=1}^2 \frac{(-1)^k }{k^2} x^2 K_2\left(k x\right), \label{4-8}
\end{align}
which is a good approximation both in high and low temperature regimes \cite{Mohamadnejad:2019vzg}. We also consider resummed daisy graphs contribution given by \cite{Carrington:1991hz}
\begin{equation}
V_{\text{daisy}} (\varphi,T) = \sum_{k=1}^{n} \frac{g_{k} T^4}{12 \pi} \left(  \left( \frac{M_k}{\nu} \frac{\varphi}{T} \right)^{3} - \left(\left( \frac{M_k}{\nu} \frac{\varphi}{T} \right)^{2}  +  \frac{\Pi_{k}(T)}{T^2} \right)^{3/2} \right) , \label{4-9}
\end{equation}
where the sum runs only over scalar bosons and longitudinal degrees of freedom of the gauge bosons \footnote{For the magnitude of theoretical uncertainties in perturbative calculations of fist-order phase transitions, including usual daisy-resummed approach see \cite{Croon:2020cgk,Gould:2021oba}}. The thermal masses
in (\ref{4-9}), $ \Pi_{k}(T) $, are given by
\begin{align}
 & \Pi_{h/\varphi} =  \frac{T^2}{24} \left(
\begin{array}{cc}
  \frac{1}{2}(9 g_{\text{SM}}^{2}+3 g_{\text{SM}}'^{2})+6 \lambda _t^2 + \lambda _{H \phi} +6 \lambda _H+ \lambda _{HS} & 0 \\
 0 & 6 g^2+\lambda _{H \phi }+6 \lambda _{\phi }+\lambda _{\phi S} \\
\end{array}
\right) , \nonumber \\
&\Pi_{S} = \frac{T^2}{24} \left(\lambda _{HS}+6 \lambda _S +\lambda _{\phi S}\right) ,  \quad \Pi_{V} = \frac{2}{3} g^{2} T^2,
 \nonumber \\
&\Pi_{W} = \frac{11}{6}  g_{\text{SM}}^{2} T^2, \quad  \Pi_{Z/\gamma}=\frac{11}{6}
\begin{pmatrix} 
 g_{\text{SM}}^{2} & 0 \\
0 &   g_{\text{SM}}'^{2} 
\end{pmatrix}
T^2 .
\label{4-10}
\end{align}

Finally, our effective potential contains Gildener-Weinberg term (\ref{4-5}) and finite-temperature contributions (\ref{4-6}) and (\ref{4-9}):
\begin{equation}
 V_{eff} (\varphi,T)  = V^{1-loop} (\varphi) + V_{T\neq0}^{1-loop} (\varphi,T) + V_{\text{daisy}} (\varphi,T). \label{4-11}
\end{equation}

In our calculations, in order to get $ V_{eff} (0,T)=0 $ at all temperatures, we subtract a constant term from potential: $ V_{eff} (\varphi,T) \rightarrow V_{eff} (\varphi,T) - V_{eff} (0,T) $.
Having potential~(\ref{4-11}), now we are ready to study phase transition.

\subsection{First order phase transition and gravitational waves} \label{4.2}
The first order phase transitions in the early Universe leave imprints in GWs which could be detected in the future.
Many beyond Standard Models predict a first-order phase transition
at the electroweak scale. This transition also provides an explanation for the matter-antimatter asymmetry in our Universe.
In the first order phase transition, just below the
critical temperature, the Universe goes from a metastable false vacuum
into a stable true vacuum, through a process of bubble nucleation, growth, and merger. Such a first-order phase transition may occur in the early Universe and naturally produces GWs \cite{Witten:1980ez,Guth:1981uk,Steinhardt:1980wx,Steinhardt:1981ct,Witten:1984rs}.
In the following we will study the dynamics of first-order phase transition and search for the parameter points of our model
that can cause such transitions.

The effective potential~(\ref{4-11}), at some
critical temperature $ T_C $, have two degenerate minima separated by a high barrier: one in $  \varphi = 0 $ and the other in $   \varphi  =\nu_C \neq 0 $:

\begin{align}
&V_{eff}(0,T_{c}) = V_{eff}(\nu_{c},T_{c}) , \nonumber \\
&\frac{d V_{eff} (\varphi,T_C)}{d \varphi} \bigg\rvert_{ \varphi = \nu_C} = 0 .
 \label{4-12}
\end{align}
By solving these two equations, one can obtain $ \nu_C $ and $ T_C $. Although, all independent parameters of the model contribute in the effective potential, we find that daisy term is negligible compared to other terms, therefore $\lambda _{\phi S}$ and $\lambda _{S}$ are irrelevant and dynamic of the phase transition only depends on $ M_S, M_V, $ and  $ g $. In figure~\ref{crit}, we have depicted $ \nu_C $ and $ T_C $ as a function of free parameters of our model.

\begin{figure}[!htb]
\begin{center}
\centerline{\epsfig{figure=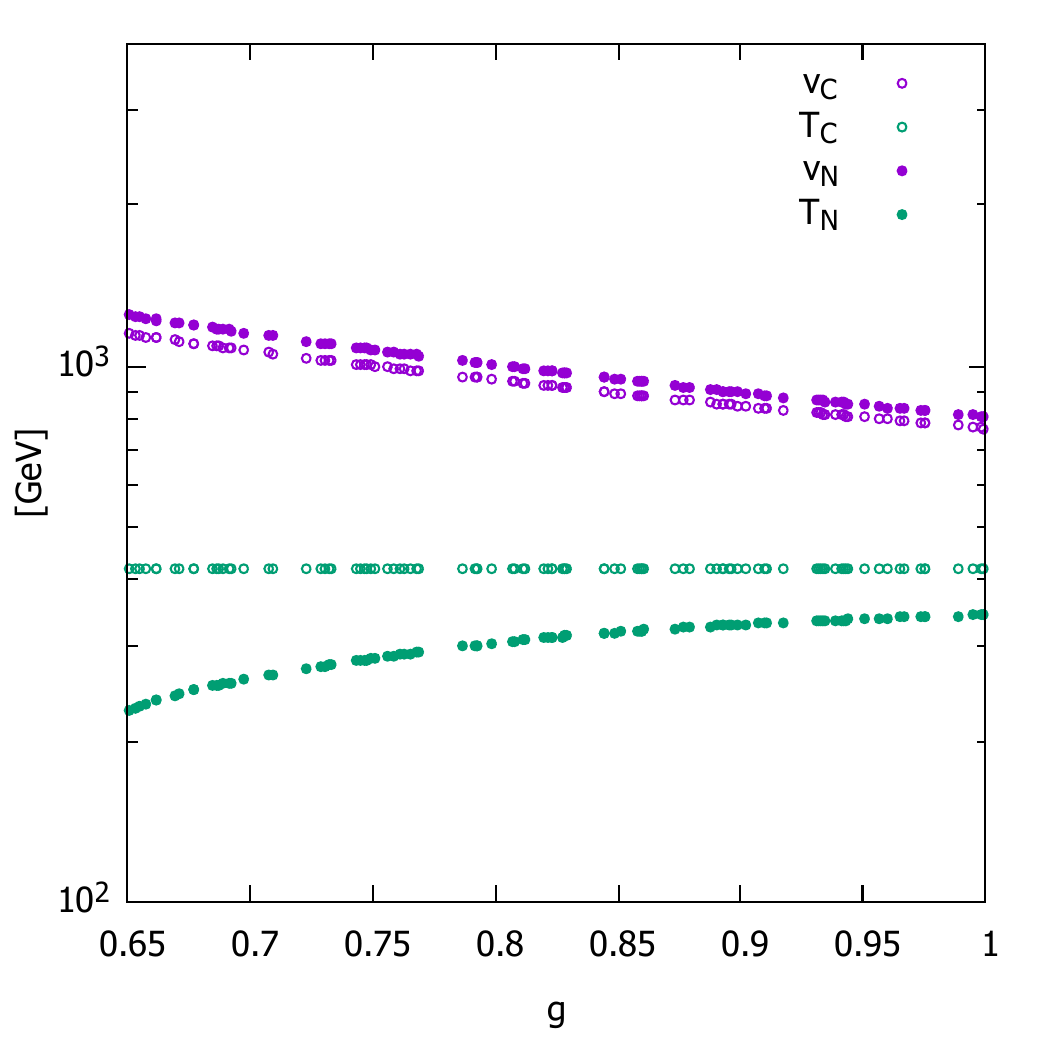,width=4.0cm}
\hspace{0.3cm}\epsfig{figure=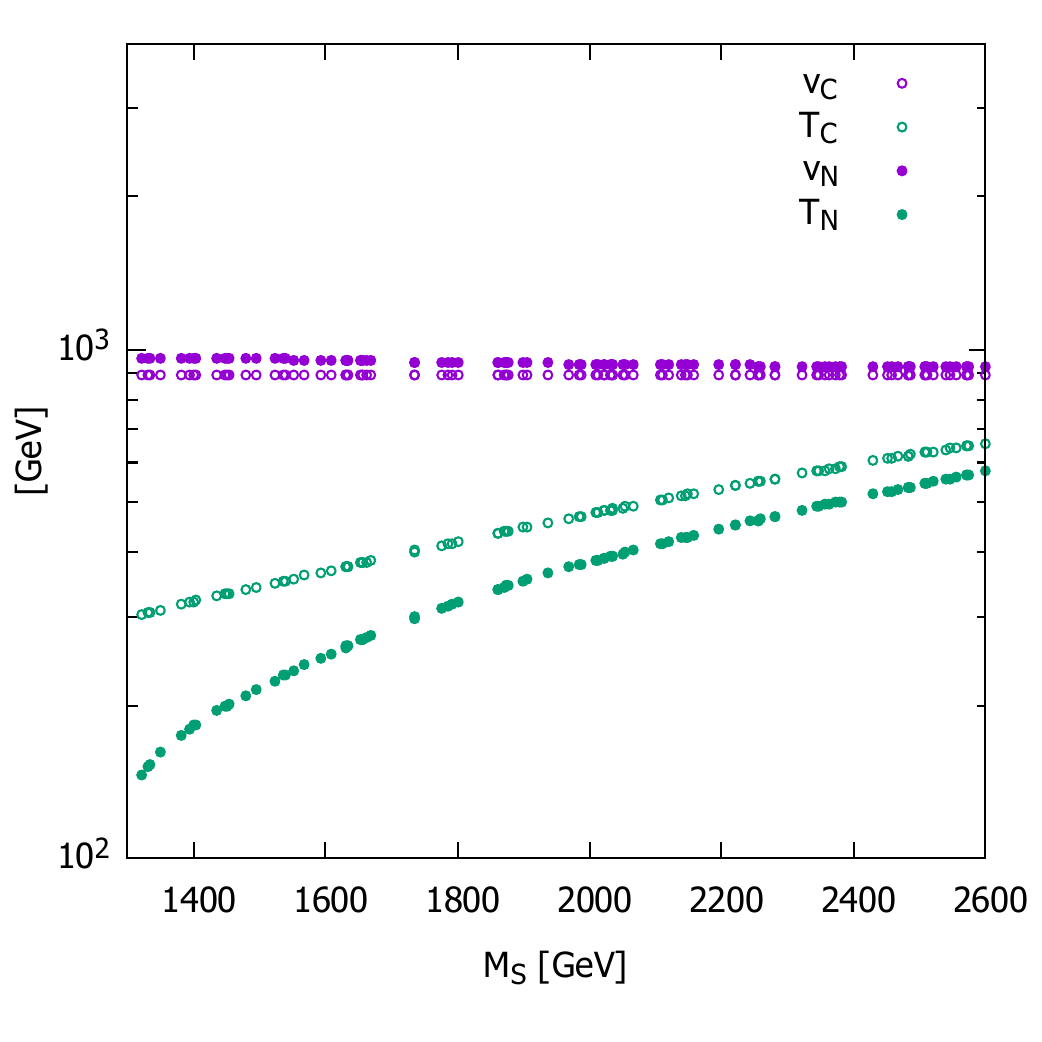,width=4.0cm}
\hspace{0.3cm}\epsfig{figure=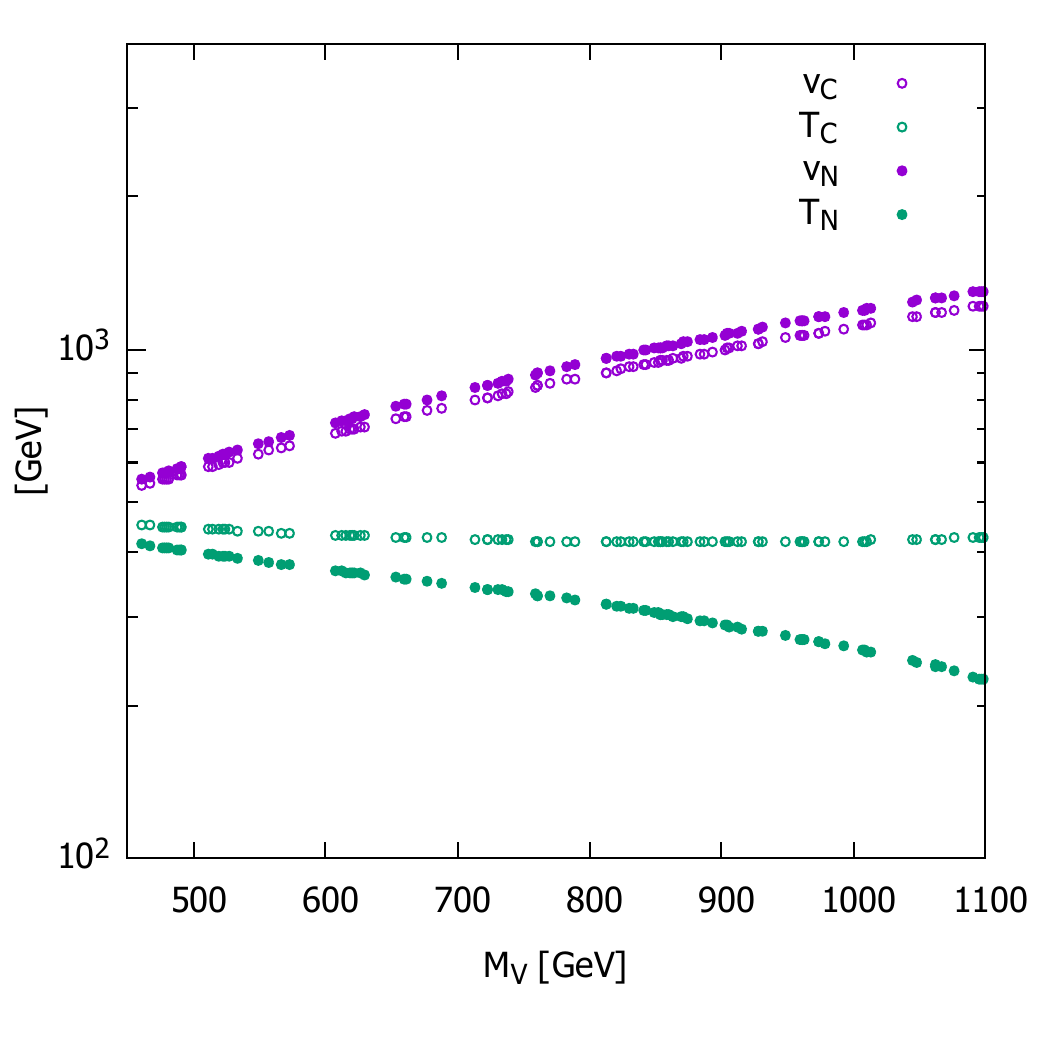,width=4.0cm}}
\centerline{\vspace{0.25cm}\hspace{0cm}(a)\hspace{4cm}(b)\hspace{4cm}(c)}
\end{center}
\vspace{-1cm}
\caption{Variation of $ \nu_C $ and $ T_C $ ($ \nu_N $ and $ T_N $) respect to parameter space. In all digrams the fixed parameters are: $ M_S = 1800 \, \text{GeV} \, , \, M_V=800 \, \text{GeV} \, , \, g= 0.85$. }\label{crit}
\end{figure}

Above the critical temperature $  \varphi  = 0 $ is the true vacuum and the symmetry is not broken. As the Universe cools down, the temperature drops below the critical one and we expect a phase transition from the false vacuum $ \varphi = 0 $ to the true vacuum $ \varphi \neq 0 $ via thermal tunneling at finite temperature.
Once this transition has happened, bubbles of the broken phase form in the sea of the symmetric phase and spread throughout the universe converting the false vacuum into the true one.

The bubbles formation starts after the temperature drops below $ T_C $, however it goes
sufficiently fast to fill the universe with bubbles of the new phase only at some
lower temperature, the nucleation temperature  $ T_N $, where the corresponding euclidean action is $ S_E= S_3(T_N)/T_N\sim140 $\footnote{This condition at the vacuum-dominated period shoud be treated more carefully (see, e.g., \cite{Kang:2020jeg}).}.
The function $ S_3(T) $ is the three-dimensional Euclidean action for a spherical
symmetric bubble given by

\begin{equation}
S_3(T) =  4 \pi \int_{0}^{\infty} dr \, r^2 \left( \frac{1}{2} \left( \frac{d \varphi}{dr} \right)^2 + V_{eff}(\varphi,T)  \right) ,  \label{4-13}
\end{equation}
where $ \varphi $ satisfies the differential equation which minimizes $ S_3 $:
\begin{equation}
\frac{d^2 \varphi}{dr^2} + \frac{2}{r} \frac{d \varphi}{dr}  = \frac{d  V_{eff}(\varphi,T) }{d \varphi} ,   \label{4-14}
\end{equation}
with the boundary conditions:
\begin{equation}
\frac{d \varphi}{dr} \bigg\rvert_{ r = 0} =0 , \quad \text{and} \quad \varphi (r \rightarrow \infty) = 0 .  \label{4-15}
\end{equation}
In order to solve eq.~(\ref{4-14}) and find the Euclidean action  (\ref{4-13}), we have used {\tt AnyBubble} package \cite{Masoumi:2017trx}.
In figure~\ref{crit}, we have also depicted $ \nu_N $ and $ T_N $ as a function of $ g $, $ M_S $, and $ M_V $.

The stochastic GW background produced by
strong first-order electroweak phase transitions comes from three contributions:
\begin{itemize}
\item  bubble walls collisions and shocks in the plasma,
\item sound waves to the stochastic background after bubble collisions but before expansion has
dissipated the kinetic energy in the plasma, and
\item turbulence forming after bubble collisions.
\end{itemize}
These three processes may coexist, and each one contributes to the stochastic GW background:
\begin{equation}
\Omega_{\rm GW} h^{2} \simeq \Omega_{\rm coll} h^{2} +  \Omega_{\rm sw} h^{2} +  \Omega_{\rm turb} h^{2} . \label{4-16}
\end{equation}

All of the above contributions are controlled by four thermal parameters:
\begin{itemize}
\item the nucleation temperature, $ T_N $,
\item the strength parameter which is the ratio of the free energy density difference between the true and false vacuum and
the total energy density, $ \alpha $,
\begin{equation}
\alpha= \frac{\Delta \left( V_{eff} - T \frac{\partial V_{eff}}{\partial T} \right)  \bigg \rvert_{T_N}}{\rho_{*}}  ,  \label{4-17}
\end{equation}
where $ \rho_{*} $ is 
\begin{equation}
\rho _*=\frac{\pi ^2 g_*}{30} T_N^4 , \label{4-18}
\end{equation}
\item the inverse time duration of the phase transition, $ \beta $,
\begin{equation}
\frac{\beta}{H_*}= T_N \frac{d}{d T} \left(  \frac{S_3 (T)}{T}\right) \bigg \rvert_{T_N} , \label{4-19}
\end{equation}
\item and the velocity of the bubble wall, $ v_w $,
which is anticipated to be close to 1 for the strong transitions \cite{Bodeker:2009qy}.
\end{itemize}

In figure~\ref{nucl}, we have depicted $ \alpha $ and $ \beta/H_* $ as a function of independent parameters of our model. For the chosen parameters we found $ \alpha \sim 10^{-2} - 10^{-1} $ and $ \beta/H_* \sim 10^{2} - 10^{3} $.

\begin{figure}[!htb]
\begin{center}
\centerline{\epsfig{figure=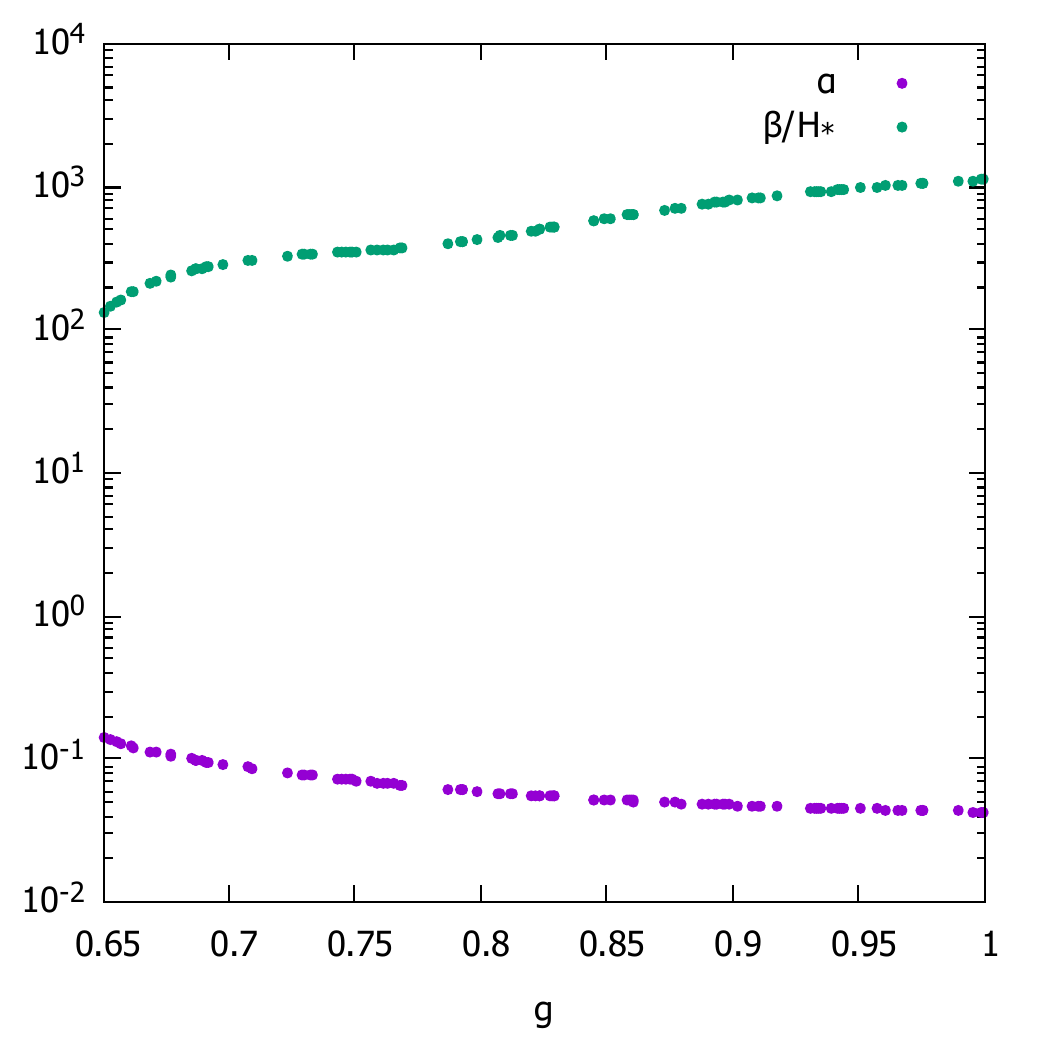,width=4.0cm}
\hspace{0.3cm}\epsfig{figure=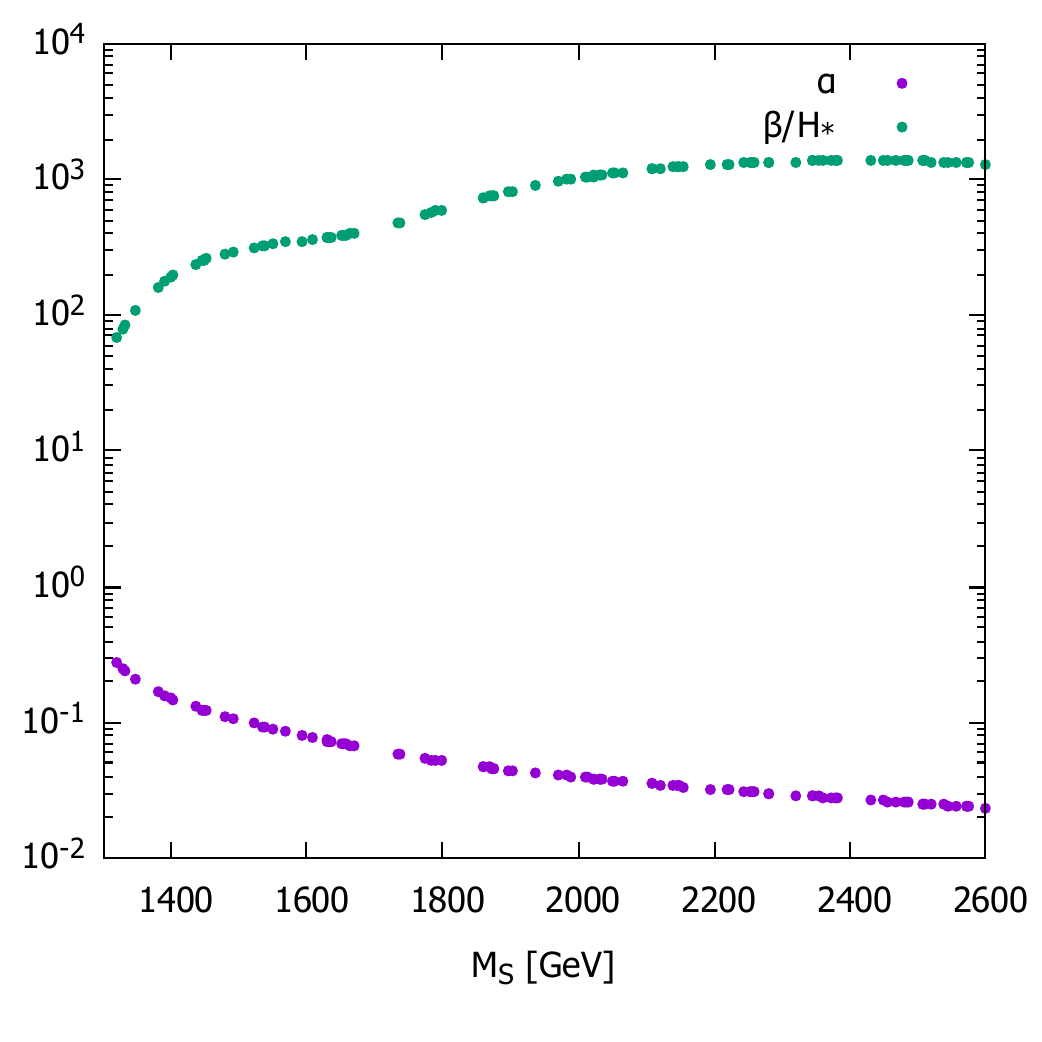,width=4.0cm}
\hspace{0.3cm}\epsfig{figure=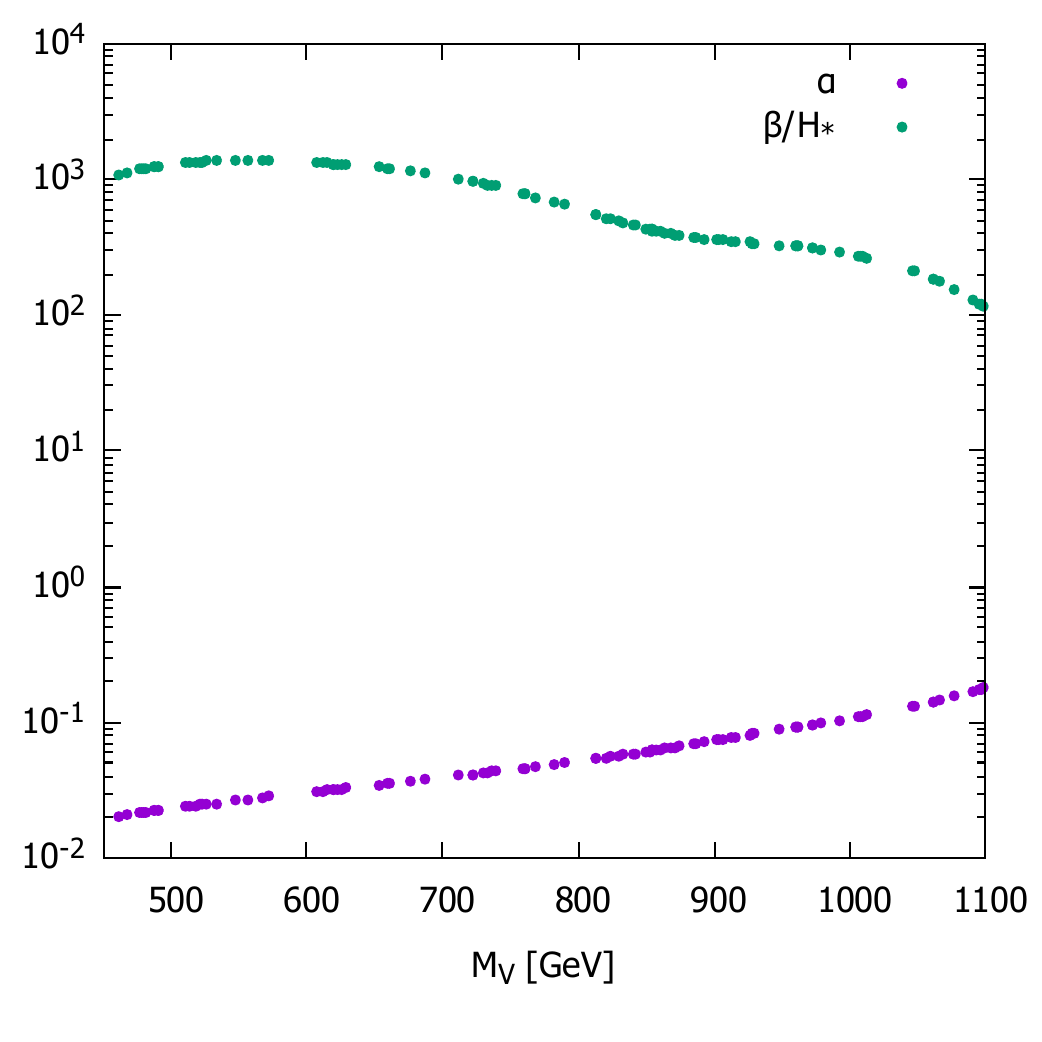,width=4.0cm}}
\centerline{\vspace{0.25cm}\hspace{0cm}(a)\hspace{4cm}(b)\hspace{4cm}(c)}
\end{center}
\vspace{-1cm}
\caption{Variation of $ \alpha $ and $ \beta/H_* $ respect to parameter space. In all digrams the fixed parameters are: $ M_S = 1800 \, \text{GeV} \, , \, M_V=800 \, \text{GeV} \, , \, g= 0.85$. }\label{nucl}
\end{figure}

In the process of the GW production, first the bubbles of the stable phase collide and merge. This stage is subdominant compared to the subsequent stages of GW
production, unless the bubbles grow as large as the Hubble length itself.
The bubble collision contribution is given by \cite{Huber:2008hg}
\begin{align}
\Omega_{\rm coll}(f) h^2= 1.67 \times 10^{-5}   \, \left( \frac{\beta}{H_*} \right)^{-2} \left( \frac{\kappa \alpha}{1+\alpha} \right)^2  
 \left( \frac{g_*}{100} \right)^{-\frac{1}{3}} \left(\frac{0.11\,v_w^3}{0.42+v_w^2}\right) \, S_{\rm coll} , \label{4-20}
\end{align}
where $ S_{\rm coll} $ parametrises the spectral shape given by
\begin{equation}
S_{\rm coll}=  \frac{3.8 \left(f/f_{\text{coll}}\right)^{2.8}}{2.8 \left(f/f_{\text{coll}}\right)^{3.8}+1} ,
\label{4-21}
\end{equation}
with
\begin{align}
f_{\text{coll}}= 1.65 \times 10^{-5} \left( \frac{ 0.62 }{ v_w^2-0.1 v_w+1.8} \right)   \left( \frac{\beta}{H_*} \right) \left( \frac{ T_N}{100} \right) \left( \frac{g_*}{100} \right)^{1/6} \, \text{Hz} .
\label{4-22}
\end{align}

After the bubble collision, the shells of fluid kinetic energy continue to expand into the plasma as sound waves. These different waves overlap and source
the dominant contribution to the GW signal which is given by\footnote{A recent study in \cite{Guo:2020grp} suggests the existence of a suppression factor for the sound wave contribution due to the finite lifetime of the GWs. The factor takes an asymptotic value of 1 for a very long 
lifetime.} \cite{Hindmarsh:2015qta}
 \begin{align}
\Omega_{\rm sw}(f) h^2= 2.65 \times 10^{-6}   \, \left( \frac{\beta}{H_*} \right)^{-1} \left( \frac{\kappa_{v} \alpha}{1+\alpha} \right)^2  
 \left( \frac{g_*}{100} \right)^{-\frac{1}{3}} v_{w} \, S_{\rm sw}. \label{4-23}
\end{align}
The spectral shape of $ S_{\rm sw} $ is
\begin{equation}
S_{\rm sw}=  \left(f/f_{\text{sw}}\right)^3 \left(\frac{7}{3  \left(f/f_{\text{sw}}\right)^2+4}\right)^{3.5} ,
\label{4-24}
\end{equation}
where
\begin{align}
f_{\text{sw}}= 1.9 \times 10^{-5} \frac{1}{v_w}
 \left( \frac{\beta}{H_*} \right) \left( \frac{ T_N}{100} \right) \left( \frac{g_*}{100} \right)^{1/6} \, \text{Hz} .
\label{4-25}
\end{align}

Finally, the last stage is the turbulent phase which its contribution to the GW spectrum is given by \cite{Caprini:2009yp}
 \begin{align}
\Omega_{\rm turb}(f) h^2= 3.35 \times 10^{-4}   \, \left( \frac{\beta}{H_*} \right)^{-1} \left( \frac{\kappa_{\rm turb} \alpha}{1+\alpha} \right)^{3/2}  
 \left( \frac{g_*}{100} \right)^{-\frac{1}{3}} v_{w} \, S_{\rm turb} , \label{4-26}
\end{align}
where
\begin{equation}
S_{\rm turb}=  \frac{\left(f/f_{\text{turb}}\right)^3}{\left(1+8 \pi  f/h_*\right) \left(1+f/f_{\text{turb}}\right){}^{11/3}} ,
\label{4-27}
\end{equation}
and
\begin{align}
f_{\text{turb}}= 2.27 \times 10^{-5} \frac{1}{v_w}
 \left( \frac{\beta}{H_*} \right) \left( \frac{ T_N}{100} \right) \left( \frac{g_*}{100} \right)^{1/6} \, \text{Hz} .
\label{4-28}
\end{align}

In eq.~(\ref{4-27}), the parameter $ h_* $ is the value of the inverse Hubble time at GW production, redshifted to today,
\begin{equation}
 h_* = 1.65 \times 10^{-5}  \left( \frac{ T_N}{100} \right) \left( \frac{g_*}{100} \right)^{1/6} .
 \label{4-29}
\end{equation} 

In the formulas of GW spectrum we have used \cite{Caprini:2015zlo,Kamionkowski:1993fg}
\begin{align}
& \kappa = \frac{1}{1+0.715 \, \alpha} (0.715 \, \alpha +\frac{4}{27} \sqrt{\frac{3 \alpha }{2}}) , \nonumber \\
&  \kappa_{v} = \frac{\alpha}{0.73 +0.083 \sqrt{\alpha }+\alpha} , \quad \kappa_{\rm turb} = 0.05 \kappa_{v},
 \label{4-30}
\end{align}
where the parameters $ \kappa $, $ \kappa_{v} $, and $ \kappa_{\rm turb} $ denote the fraction of latent heat that is transformed into gradient energy of the Higgs-like field, bulk
motion of the fluid, and MHD turbulence, respectively.

\section{Results} \label{sec5}
After having studied DM phenomenology and electroweak phase transition of our two-component DM model in
the previous sections, we will now concentrate on the question of
whether it is feasible to correctly reproduce the known
features of DM and first order phase transition at the same time.
On the DM side, we have relic density constraint as well as the upper bound of DM-Nucleon cross section obtained in direct detection experiments. On the other hand, we are looking for strong first order phase transition which leads to GW production at the early Universe.
Obviously, the above requirements will
impose constraints on the parameter space of the model,
 which is the subject of the present
section.

In order to obtain the parameter space consistent with DM relic density (see eq.~(\ref{3-6})) and direct detection constraint (see eq.~(\ref{3-15})), we should scan over four independent parameters of the model, i.e., 
$ M_S, M_V, g, \, \text{and} \, \, \lambda _{\phi S}$. To do so, regarding the strong constraint on DM-Nucleon cross section, we first obtain $ \overline{g} $ and $ \overline{\lambda}$ using the following equations
\begin{equation} \label{5-1}
\alpha_v \bigg\rvert_{g = \overline{g}} = 0, \quad  \alpha_s \bigg\rvert_{\lambda _{\phi S}=\overline{\lambda}} = 0.
\end{equation}
Considering eqs.~(\ref{2-7}), (\ref{2-8}), (\ref{3-8}) and (\ref{3-12}), the solutions are given by
\begin{align}  \label{5-2}
\overline{g}(M_V,M_S) &= \frac{M_V}{\sqrt{\frac{\sum_{k=1}^{n} g_{k}  M_{k}^{4}}{8 \pi^2 M_h^2}-\nu_1^2}}. \nonumber \\
\overline{\lambda}(M_V,M_S,g) &= \frac{2 M_S^2 \left(\sin^2\alpha M_h^2+\cos^2\alpha M_{\varphi }^2\right)}{\nu^2 \cos^2\alpha M_{\varphi }^2}.
\end{align}
Now, looking for the correct value of DM relic density (\ref{3-6}) and regarding perturbativity constraints (all couplings $ < 4 \pi $)), we scan over random values of $ M_S $ and $ M_V $, while we choose $ 0.9 \, \overline{g} < g < 1.1 \, \overline{g} $ and $ 0.9 \, \overline{\lambda} < \lambda _{\phi S} < 1.1 \, \overline{\lambda} $.
In this way, according to eq.~(\ref{3-13}), we restrict ourselves to the small values of $ \sigma_s $ and $ \sigma_v $ which can possibly evade the direct detection constraint (\ref{3-15}).
According to this strategy, we obtain DM relic density for $ 40 \, \text{GeV} \lesssim M_{DM} \lesssim 2000 \, \text{GeV} $.

\begin{figure}[!htb]
\begin{center}
\centerline{\hspace{0cm}\epsfig{figure=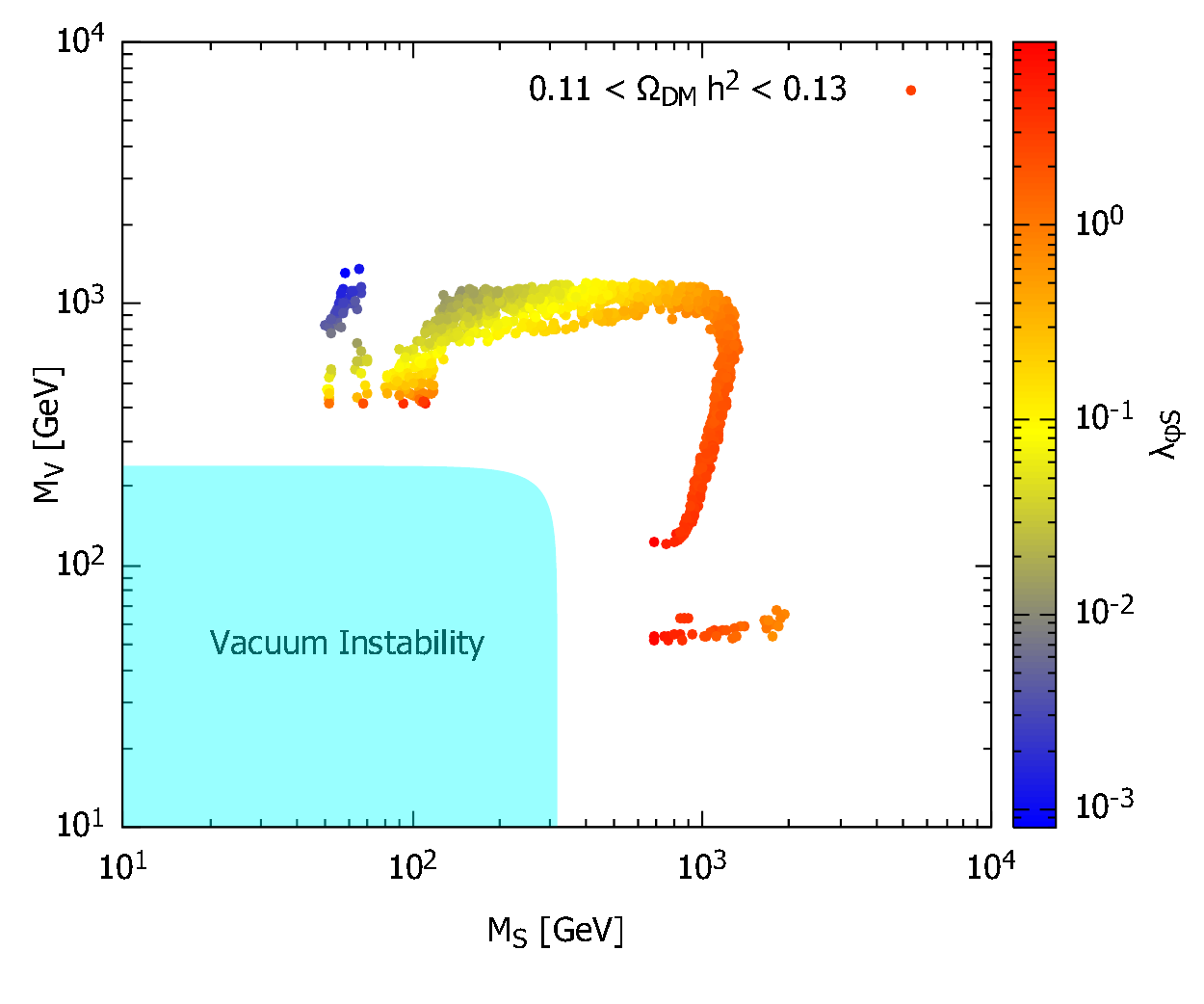,width=6cm}\hspace{0.3cm}\epsfig{figure=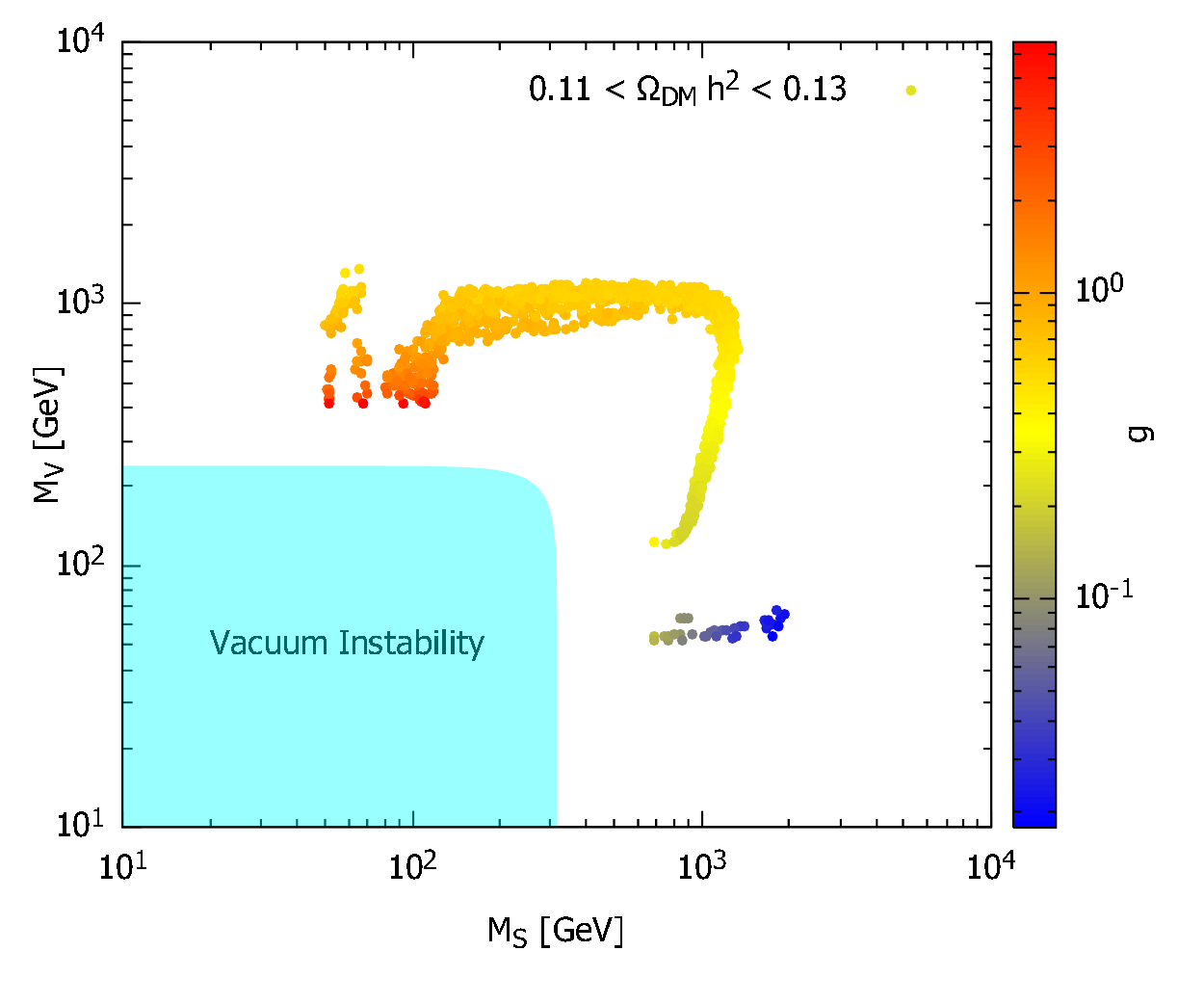,width=6cm}}
\centerline{\vspace{0.5cm}\hspace{0.5cm}(a)\hspace{6cm}(b)}
\centerline{\hspace{0cm}\epsfig{figure=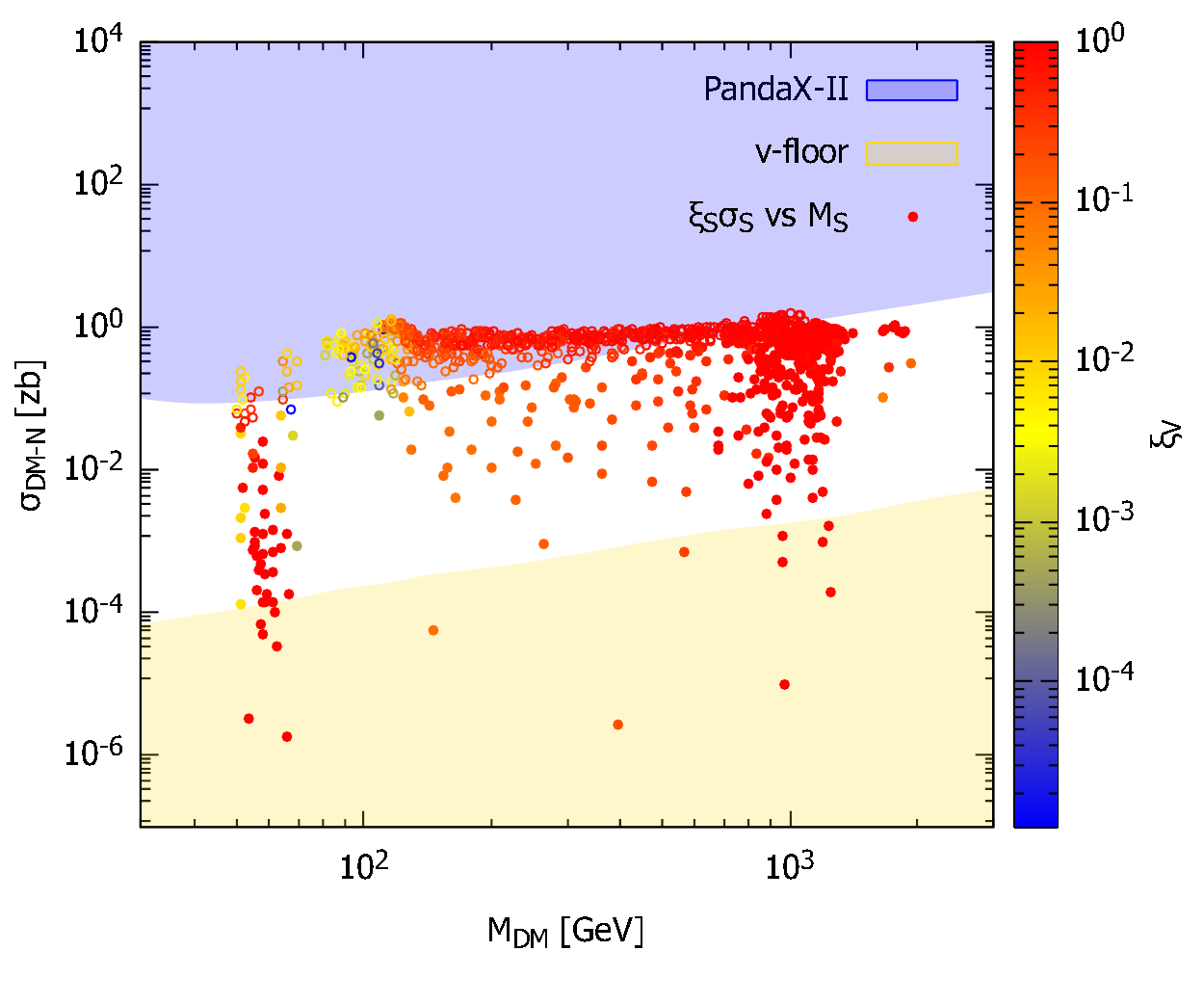,width=6cm}\hspace{0.3cm}\epsfig{figure=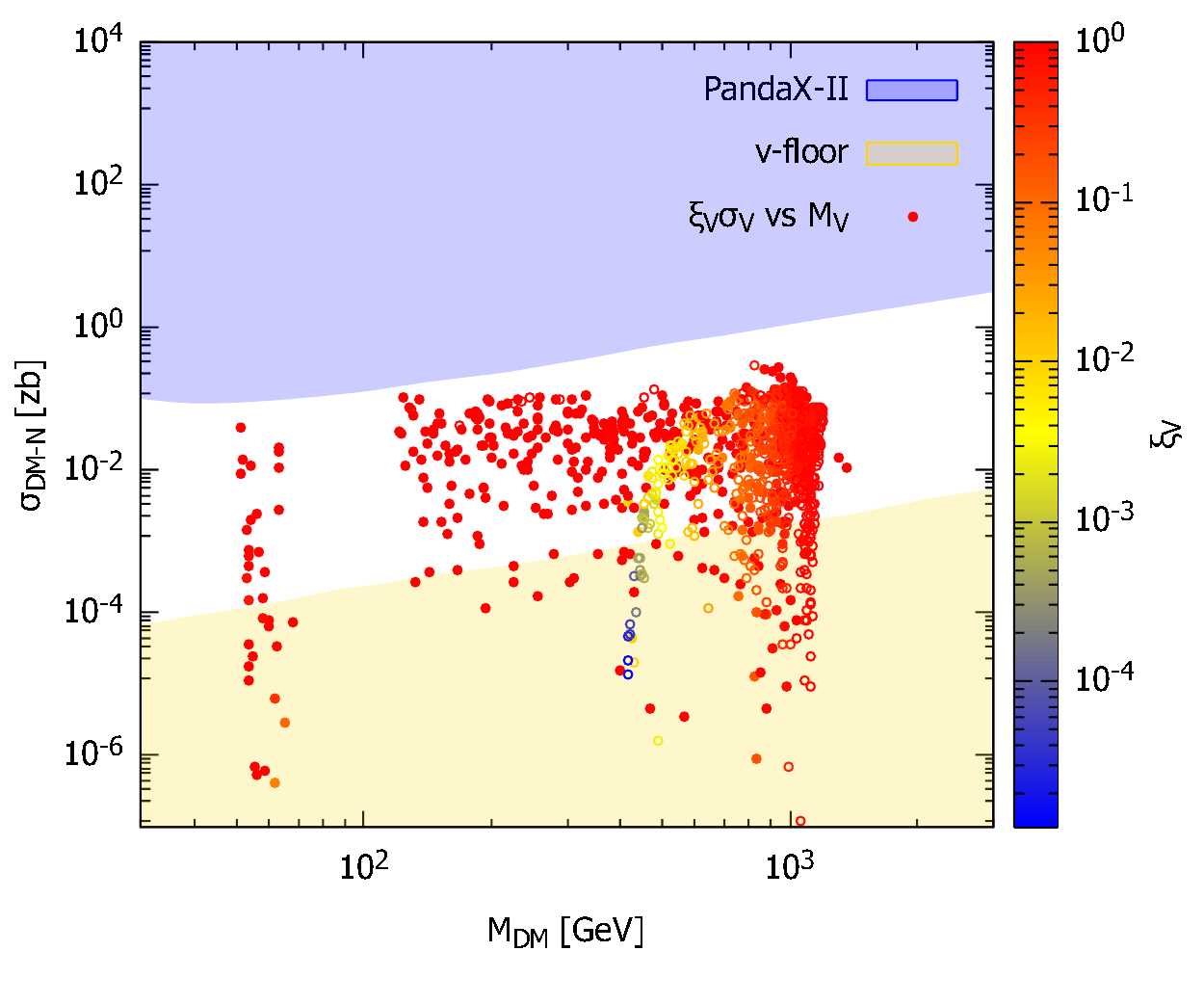,width=6cm}}
\centerline{\vspace{-1.2cm}\hspace{0.5cm}(c)\hspace{6cm}(d)}
\centerline{\vspace{-0.0cm}}
\end{center}
\caption{(a) and (b): parameter space consistent with DM Relic density. Cyan region is not allowed because in this region $ M_{\varphi}^2 < 0 $ (see eq.~{\ref{2-7}}) and we do not have a global vacuum. (c) and (d):  Rescaled DM-Nucleon cross section for the parameter space already constraind by DM relic density. Hollow circles are excluded by direct detection constraint (\ref{3-15}).}\label{dark}
\end{figure}

In figure \ref{dark} (a) and (b), the parameter space consistent with DM relic density is obtained based on our strategy. In (c) and (d), for these parameters, rescaled DM-Nucleon cross sections, i.e., $ \xi_{S} \sigma_s $ and $ \xi_{V} \sigma_v $, are also depicted. As we see, there are some points between the PandaX-II direct detection bound and the neutrino floor which can be probed in the future direct detection experiments. Although DM relic density and direct detection experiments restrict the model, there are some parts of the parameter space which is not excluded yet. In 
figure \ref{dark} (c) and (d), we have also depicted the fraction of vector DM, $ \xi_{V} $. As we see, for some scatter points, vector DM is dominated, while for the other points DM mostly consists of scalar DM. Note that, although our strategy strongly restricts the parameter space in order to get unconstrained DM-Nucleon cross section, still there are some points which violate direct detection constraint (\ref{3-15}).
This points are depicted with hollow circles and they should be excluded even when they are below the upper bound of Pandax-II. Therefore, considering $ \xi_{S} \sigma_s $ and $ \xi_{V} \sigma_v $ separately is not enough and in order to constrain two component DM models by direct detection experiments, one should combine both signatures.

Now looking for the first order electroweak phase transition and GW, we scan over the parameter space once more.
In accordance with the DM results, we choose the same range of the parameters as figure \ref{dark}. The result is shown in figure \ref{grav}.

\begin{figure}[!htb]
\begin{center}
\centerline{\epsfig{figure=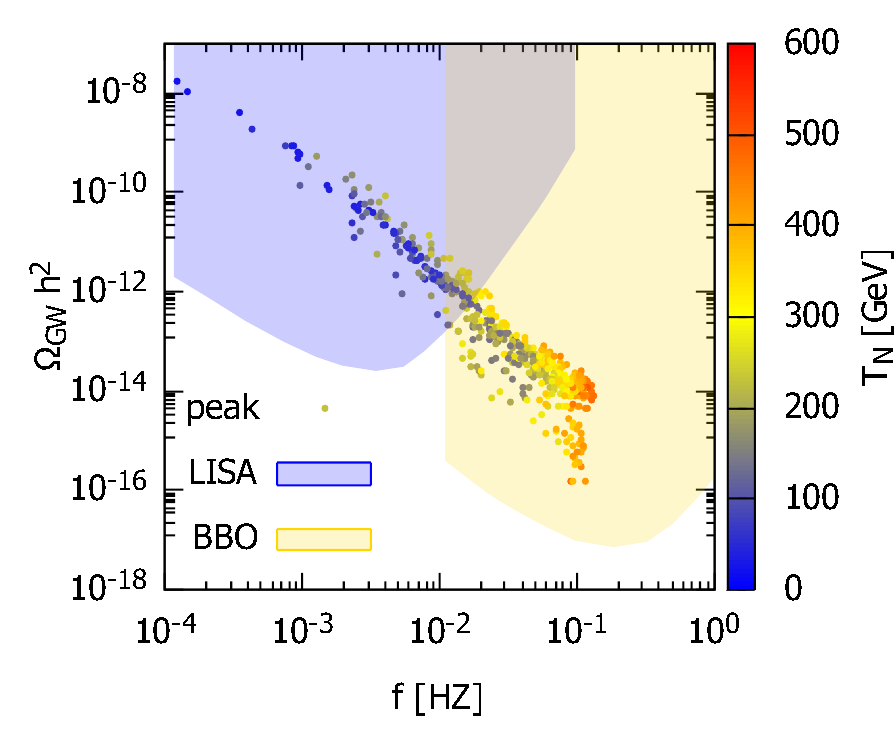,width=4.6cm}
\hspace{0.0cm}\epsfig{figure=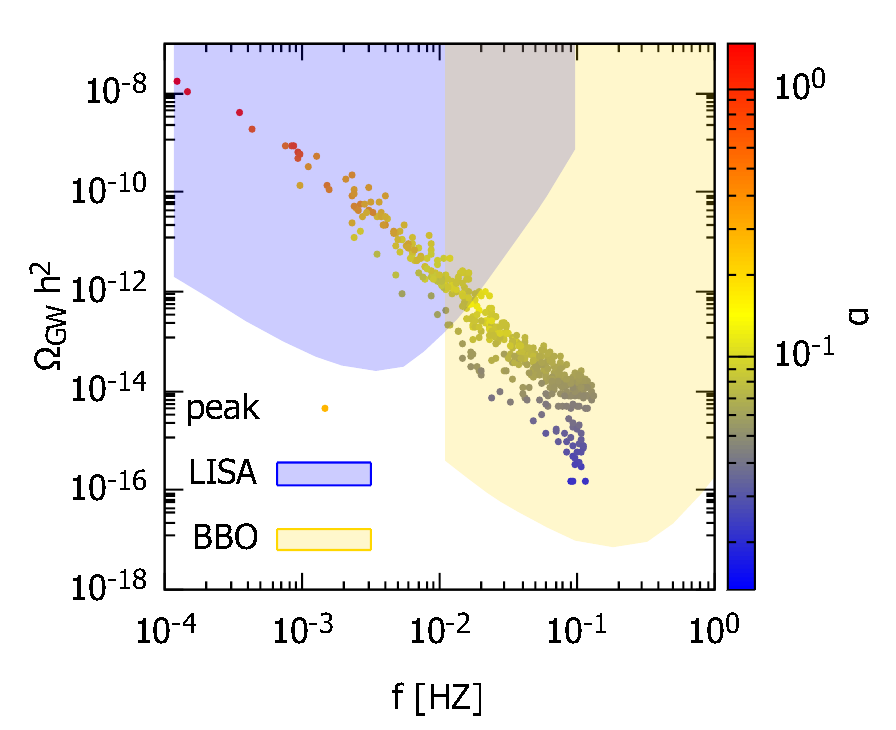,width=4.6cm}
\hspace{0.0cm}\epsfig{figure=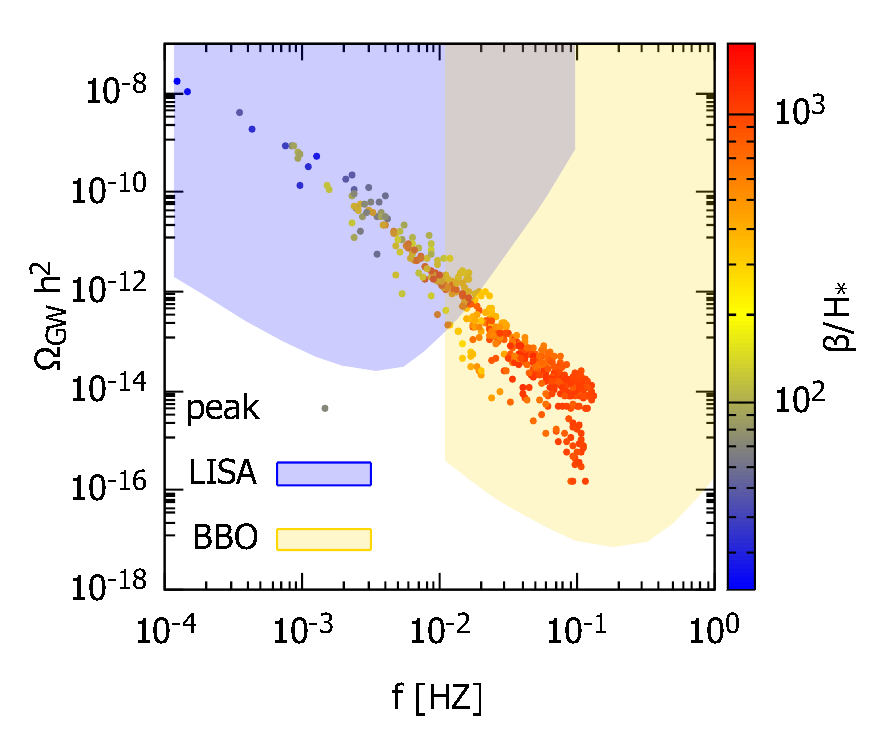,width=4.6cm}}
\centerline{\vspace{0.25cm}\hspace{0cm}(a)\hspace{4cm}(b)\hspace{4cm}(c)}
\end{center}
\vspace{-1cm}
\caption{The peak of GW spectrum against frecuency. LISA and BBO sensitivities are also depicted.}\label{grav}
\end{figure}

In study of phase transition and GW, the relevant parameters are  $ M_S, M_V, \, \text{and} \, \, g$. We found that for around 14 percent of the parameter space scanned here, first order phase transition can occur, generating a stochastic background of GWs. Most peaks of the GW spectrum are detectable by LISA and BBO detectors\footnote{For a new type of sensitivity curves for gravitational-wave signals from cosmological first order phase transitions for LISA and BBO see \cite{Alanne:2019bsm,Schmitz:2020syl,Schmitz:2020rag}}.

So far, we have studied DM phenomenology and phase transition separately. Now we consider both aspects simultaneously which gives us figure \ref{darkgrav}. In this figure, we have used the scatter points already obtained in the study of DM phenomenology, and saw if they also produce first order phase transition and GWs.
In general, the DM freeze-out temperature $ T_{F} \sim M_{DM}/20 $ may be greater than nucleation temperature $ T_{N} $ in some parameter points.
In this case, as the phase transition is not completed, the freeze-out can be affected. For phenomenology of a late phase transition see \cite{Cohen:2008nb}.
However, for the parameter points in figure \ref{darkgrav}, we have compared freeze-out temperature with nucleation temperature and find out $ T_{F} < T_{N} $ where $ T_{F} \sim max(M_S,M_V)/20 $. Therefore, this issue does not affect our result and the DM properties would not be modified between $ T_{F} $ and the present day, at least for the parameter space considered in figure \ref{darkgrav} (where $ 0.31 \lesssim T_F / T_N \lesssim 0.47 $).

\begin{figure}[!htb]
\begin{center}
\centerline{\hspace{0cm}\epsfig{figure=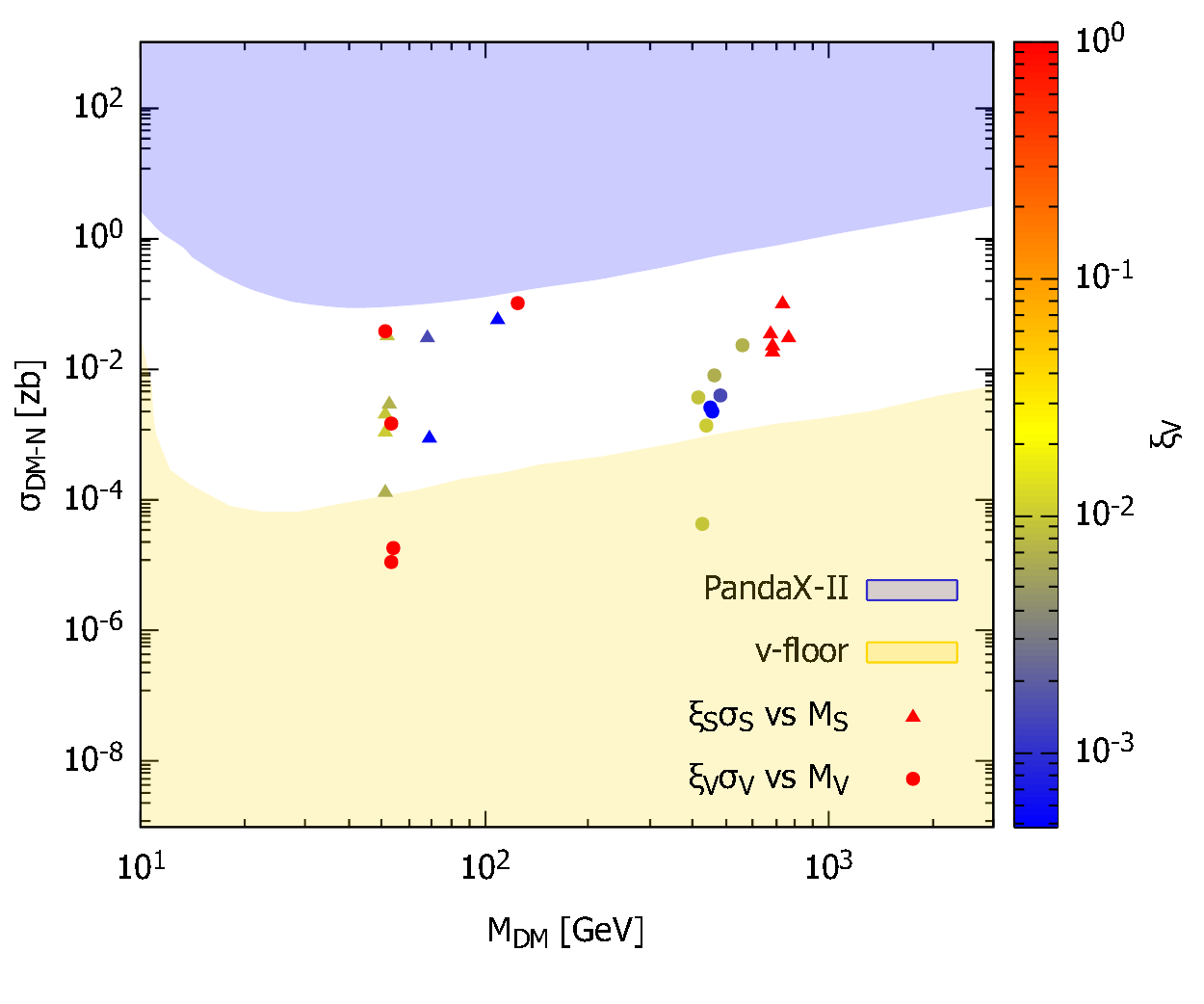,width=6cm}\hspace{0.3cm}\epsfig{figure=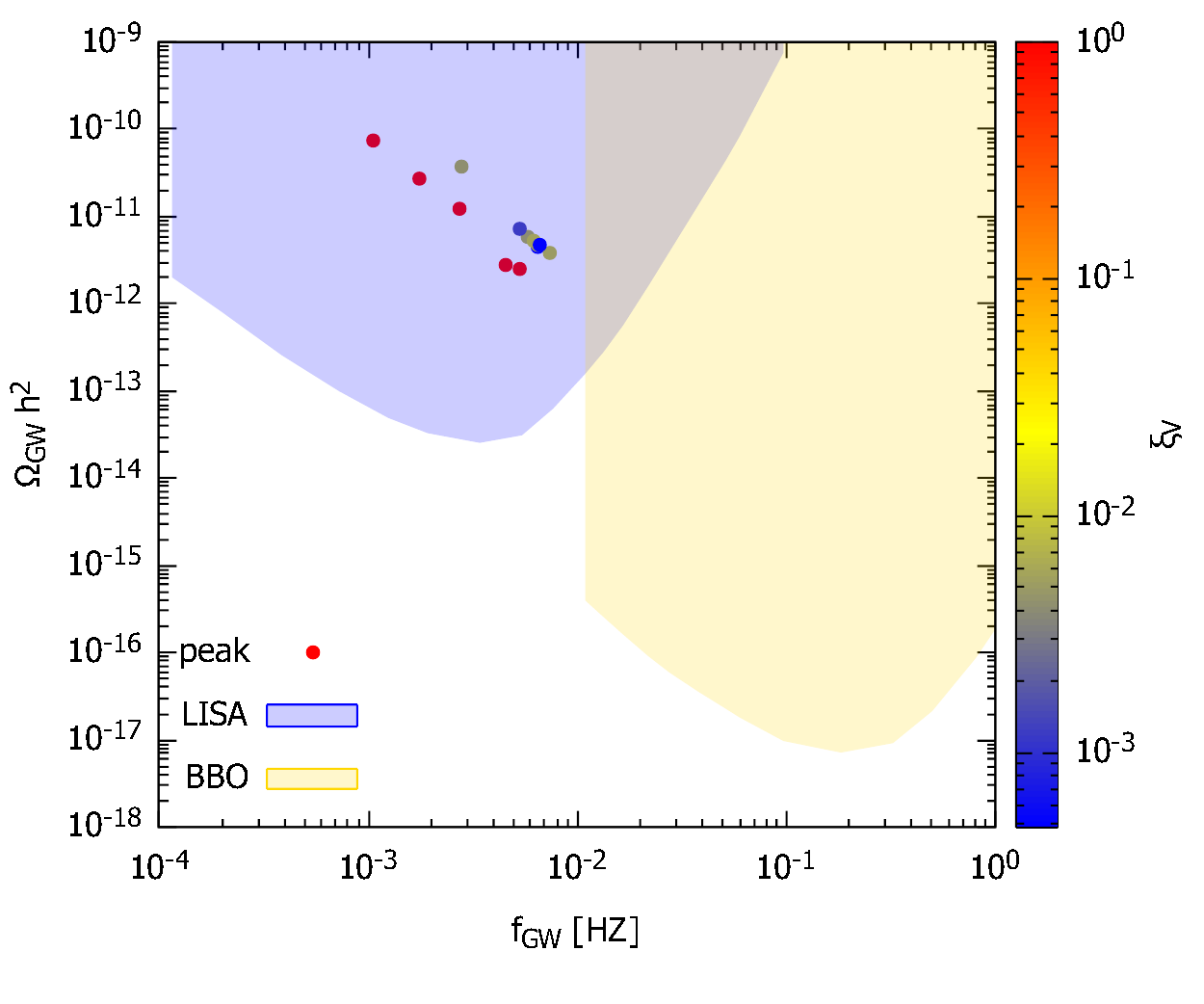,width=6cm}}
\centerline{\vspace{0.5cm}\hspace{0.5cm}(a)\hspace{6cm}(b)}
\centerline{\vspace{-1.5cm}}
\end{center}
\caption{(a): Rescaled DM-Nucleon cross section VS DM mass. The scatter points are already consistent with DM relic density (\ref{3-6}) and direct detection (\ref{3-15}) constraints. (b): GW peak VS frequncy for the same points of the parameter space.}\label{darkgrav}
\end{figure}

As figure \ref{darkgrav} implies, for some points in the parameter space, the model is consistent with DM constraints, while at the same time generates first order electroweak phase transition and GWs detectable by LISA and BBO. To be more explicit, we have chosen two benchmark points given in table \ref{table1}. In this table all relevant quantities, including independent parameters of the model, DM properties, and phase transition parameters, are given.

\begin{table}
\centering
\begin{tabular}{c c c c c c} 
 \hline
 \hline
$ \# $ & $ \lambda_{\phi S} $ & $ g $ & $ M_{S} $ (GeV) & $ M_{V} $ (GeV) & $ M_{\varphi} $ (GeV) \\ [0.5ex] 
 \hline
1 & 0.036 & 1.469 & 52.37 & 559.4 & 132.3 \\
2 & 5.491 & 0.118 & 761.0 & 53.78 & 124.4 \\
 [1.5ex]
 \hline
 \hline
$ \# $ & $ \Omega_S h^2 $ & $ \Omega_V h^2 $ & $ \Omega_{DM} h^2 $ & $ \xi_{S} \sigma_S $ (zb) & $ \xi_{V} \sigma_V $ (zb) \\ [0.5ex] 
 \hline
1 & $1.09 \times 10^{-1}$ & $7.72 \times 10^{-4}$ & $1.10 \times 10^{-1}$ & $2.93 \times 10^{-3}$ & $2.29 \times 10^{-2}$ \\
2 & $1.00 \times 10^{-3}$ & $1.12 \times 10^{-1}$ & $1.13 \times 10^{-1}$ & $2.99 \times 10^{-2}$ & $1.83 \times 10^{-5}$ \\
 [1.5ex]
 \hline
 \hline
$ \# $ & $ T_C $ (GeV) & $ T_N $ (GeV) & $ \alpha $ & $ \beta/{H_*} $ & $ (\Omega_{GW} h^2)_{max} $ \\ [0.5ex] 
 \hline
1 & 135.9 & 78.19 & 0.257 & 187.2 & $3.71 \times 10^{-11}$ \\
2 & 161.0 & 81.14 & 0.234 & 68.72 & $7.59 \times 10^{-11}$ \\
\hline
\end{tabular}
\caption{Two benchmark points with DM and phase transition parameters.}
\label{table1}
\end{table}

For these benchmark points, the GW spectrum is depicted in figure \ref{spectrum}. For both benchmark points, around the peak, the dominated contribution of GW signal is sound wave. For the first benchmark point, scalar DM is dominant, while for the second one, vector DM makes most of DM relic density. The peak of the GW spectrum for both benchmark points falls within the observational window of LISA.

\begin{figure}[!htb]
\begin{center}
\includegraphics[scale=0.6]{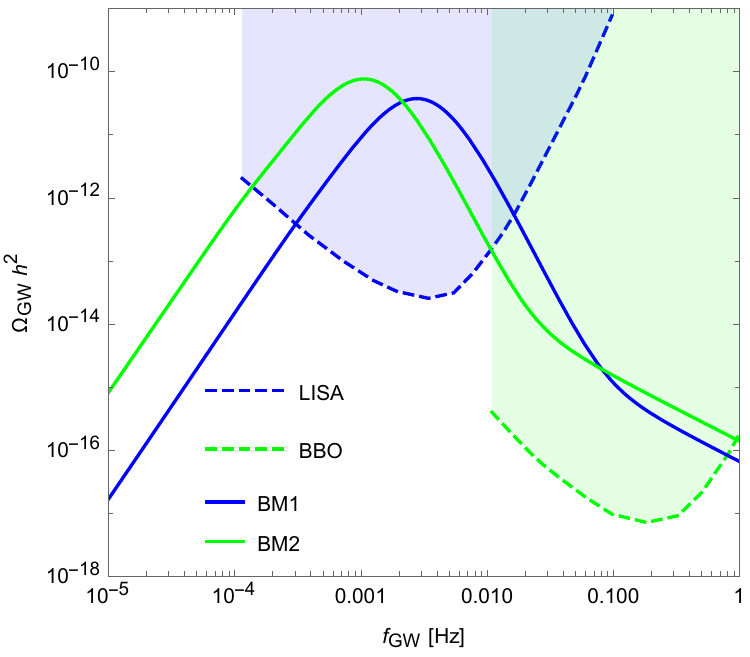}
\end{center}
\vspace{-0.5cm}
\caption{GW spectrum for benchmark points of the table \ref{table1}. \label{spectrum}}
\end{figure}

Finally, we should mention that sound wave contribution dominates for all scatter points in figure \ref{darkgrav} and it is almost indistinguishable from the sum of the three sources, at least around the GW peak. However, far from the peak, this is not necessarily the case. For example, after the fracture of the first benchmark curve around 0.1 HZ, the bubble collision
contribution will be dominated.

\section{Conclusion} \label{sec6}
In this work we studied a two-component DM model as an extension of the SM with classical scale symmetry. It realizes electroweak symmetry breaking through Gildener-Weinberg mechanism and
gives a natural solution to the hierarchy problem.
The model consists three new fields: a real scalar, a complex scalar, and a vector field which two of them, the real scalar field and vector field, can play the role of DM.  Our two-component DM model is obtained by adding them to the SM via the Higgs-portal.
The Boltzmann equations
for both DM components were solved numerically and in order to determine a region of parameter space that is consistent with
Planck and PandaX-II data a scan over four dimensional parameter space was performed.

After introducing the
model and investigating DM phenomenology,
we focused on the phase transition dynamics.
With the aim of exploring the nature and the strength of the electroweak phase transition, the full finite-temperature effective potential of the model at one loop level has been obtained. Despite the absence of a barrier in the
zero-temperature potential, it was demonstrated that the finite-temperature effects induce a barrier between the symmetric and the broken
phase vacua, and thereby give rise to a first-order electroweak phase transition which can generate GWs.

The spectrum of these GWs can be described in terms of only four properties: the bubble nucleation temperature $ T_N $, the strength parameter
$ \alpha $, the transition rate parameter $ \beta $, and the bubble wall speed $ v_w $. These are all computable from the underlying particle physics model and therefore are functions of the independent parameters of our model. The space missions such as
LISA and BBO could detect these GWs if the phase transition took place at scale of electroweak symmetry-breaking.
LISA and BBO are particle physics experiments, as well as  astrophysical observatories. Albeit, much should be
done to realize the goal of making LISA into a particle physics experiment to complement the Large Hadron Collider (LHC). On the other hand,
although continuing efforts at
the LHC will be able to examine some of the beyond Standard Models, there remain many that cannot
be probed through collider experiments on a timescale as good as LISA and BBO, if at all.

Our results indicate the model can survive DM relic density and direct detection constraints, while at the same time produce GWs during the first order electroweak phase transition.
A positive GW signal at
LISA and BBO would most likely point toward new physics at the TeV scale such as the classically scale invariant potential studied here.



\providecommand{\href}[2]{#2}\begingroup\raggedright\endgroup

\end{document}